\documentclass[fleqn,usenatbib]{rasti}

\usepackage{newtxtext,newtxmath}

\usepackage[T1]{fontenc}

\DeclareRobustCommand{\VAN}[3]{#2}
\let\VANthebibliography\thebibliography
\def\thebibliography{\DeclareRobustCommand{\VAN}[3]{##3}\VANthebibliography}

\usepackage{graphicx}	
\usepackage{amsmath}	
\usepackage{booktabs} 

\title[aRieL - Exoplanet Observatory Scheduling]{aRieL - Reinforcement Learning for the Ariel Space Telescope Scheduling}

\author[James Kostas Ray et al.]{
James Kostas Ray,$^{1}$\thanks{E-mail: james.ray.16@ucl.ac.uk}
Kai Hou Yip,$^{2}$
Alexandra Thompson,$^{3,1}$
Lu{\'i}s F. Sim{\~o}es$^{4}$
and Nikolaos Nikolaou$^{1}$
\\\\
$^{1}$Department of Physics and Astronomy, University College London, Gower Street, London WC1E 6BT, UK\\
$^{2}$Department of Physics, King's College London, University of London, Strand, London, WC2R 2LS, United Kingdom\\
$^{3}$INAF-Osservatorio Astronomico di Palermo, Piazza del Parlamento, 1, 90134 Palermo, Italy\\
$^{4}$ML Analytics, 2660-329 Lisbon, Portugal
}

\date{Accepted XXX. Received YYY; in original form ZZZ}

\pubyear{\the\year{}}

\begin{document}
\label{firstpage}
\pagerange{\pageref{firstpage}--\pageref{lastpage}}
\maketitle

\begin{abstract}
The Ariel mission will conduct a population level survey of hundreds of exoplanet atmospheres, requiring observations to be scheduled across a large and diverse catalogue of potential targets. By framing Ariel scheduling as a long-horizon optimisation problem in which individual decisions affect both mission time and the scientific opportunities available later in the survey, Reinforcement Learning (RL) can be utilised as a framework for target selection. We introduce \texttt{aRieL}, a simulated observing environment coupled to a Set Transformer Proximal Policy Optimisation (PPO) agent that is trained to generate a scheduling policy. The RL policy consistently finds a strong compromise between competing survey objectives, producing large Tier~1 samples while maintaining high Tier~3 completion, population coverage, and observing efficiency when compared to a set of baseline heuristic polices. This behaviour is retained when the candidate catalogue is substantially expanded, while modifications to the reward function produce corresponding changes in the learned observing strategy. We further show that a policy trained on the baseline mission scenario can generalise without retraining to a modified scenario requiring repeated Tier~3 observations, revealing the resulting trade-off with the wider survey. These results demonstrate that reinforcement learning provides a flexible approach to large-scale astronomical scheduling, in which the observing strategy can respond directly to changes in mission state and to the scientific priorities, and motivate its wider exploration for observatories with similarly complex, state dependent scheduling problems.
\end{abstract}




\section{Introduction}

Ariel, ESA’s M4 Cosmic Vision mission, will deliver the first large-scale survey of exoplanet atmospheres, characterising a diverse population of hundreds of transiting planets. Through simultaneous visible and infrared spectroscopy spanning 0.5-7.8$\mu$m, Ariel aims to measure atmospheric composition and structure across a broad range of planetary and stellar properties, enabling population-level studies of the processes that govern planetary formation and evolution \citep{tinetti2020ariel}. Ariel is not simply tasked with maximising the number of completed observations. Its scientific return depends on sampling a broad population of planets across a range of planetary and stellar properties with a high enough signal to noise ratio for characterisation. The candidate population will remain larger than the final observed sample and is expected to evolve as new systems are discovered and characterised \citep{edwards2019updated, edwards2022ariel}. Target selection and scheduling are therefore coupled: deciding which observation to make affects both the remaining mission time and the scientific composition of the final survey. 

Ariel adopts a tiered observing strategy to balance population coverage against the depth of atmospheric characterisation. Targets therefore require different numbers of transit or eclipse observations depending on their assigned science tier and observational properties \citep{tinetti2020ariel,edwards2022ariel}. Moreover, the number of suitable candidate targets substantially exceeds the number that can be observed during the mission's lifetime, making the distribution of observations across planetary and stellar parameter space an important component of the survey design. Since these observations occur within discrete, time dependent transit and eclipse opportunities, constructing the final observing programme requires target selection and scheduling to be considered together. The result is an optimisation problem in which limited observing time must be allocated between competing targets while maintaining the desired scientific coverage of the population. Decisions are coupled because selecting one observation consumes time and alters subsequent opportunities. The target population and scientific priorities can also evolve. This makes fixed prioritisation rules or locally optimal decisions potentially inadequate for optimising the programme over the full mission. Existing Ariel scheduling work has approached the problem primarily through combinatorial and metaheuristic optimisation, including hybrid multi-start approaches and Forgetful Swarm Optimisation \citep{morales2022ariel,nakhjiri2023hybrid,nakhjiri2024forgetful}. These methods have demonstrated the ability to construct and iteratively improve complete mission schedules across several Ariel scheduling scenarios. Here we instead ask whether the scheduling problem can be represented by a learned sequential policy, where the next observation is selected directly from the current mission state. 

Reinforcement Learning (RL) is a Machine Learning (ML) approach for interactive decision making systems and has been explored for a range of astronomical operations, including adaptive optics control and radio interferometric calibration \citep{landman2021self,yatawatta2021deep}. Of particular relevance to observation planning, RL-based approaches have been proposed for the scheduling and control of individual telescopes and distributed telescope arrays \citep{jia2023simulation,jia2023observation, terranova2023self, puangragsa2025implementation, zhang2025solving}. For Ariel the main challenges are dealing with long episodes (being the duration of the science operations phase) whilst trying to balance multiple objectives over a long horizon. Ariel is currently planned for launch in 2031 and has a nominal four year operational lifetime. In this work we follow previous Ariel scheduling studies in adopting a conservative estimate of a 3.5 year science scheduling horizon. 

We introduce \texttt{aRieL}, a reinforcement learning environment for Ariel observation scheduling, along with a trained Set Transformer Proximal Policy Optimisation agent designed to select between dynamically changing observation opportunities. Through a series of controlled deterministic experiments, we test whether the learned policy can remain competitive across different candidate populations, preserve population diversity, respond to changes in the reward objective, and adapt to a modified Tier~3 observing requirement without retraining. This environment can be used in conjunction with a variety of methods but is mainly used to enable RL as a viable approach for the scheduling task. 

\section{Reinforcement Learning (RL)}

ML methods learn representations and patterns from data, with the objective of mapping features to a set of labels or capturing the underlying distribution and properties of the data itself. This is often performed in a static setting, where an algorithm is trained on a fixed dataset and is subsequently provided with new data to map into the learned representation space. This representation can then be used for downstream tasks such as regression, classification, feature extraction, or anomaly detection. RL follows a different design philosophy. Rather than learning from a fixed dataset as most ML methods do, an RL agent continuously interacts with an environment, with each interaction potentially changing the environment. At each step, the agent is provided with a snapshot, or state, of the environment and selects an action, which is then applied to the environment. The agent learns which actions are favourable through a task specific reward function, with the objective of maximising the cumulative reward obtained over time. For example, this reward might correspond to scoring points in a game such as \textit{Pong}, accumulating returns in a financial trading environment, or maximising the number of Ariel Tier 3 targets observed in a year. For a general introduction to reinforcement learning, we refer the reader to \citet{de2018multi} or \cite{barto2021reinforcement}.

Formally speaking, RL is described as a Markov Decision Process (MDP),
\begin{equation}
    \mathcal{M} = \left(\mathcal{S}, \mathcal{A}, P, R, \gamma\right),
\end{equation}
where $\mathcal{S}$ and $\mathcal{A}$ denote the state and action spaces, respectively,  $P(s_{t+1}\mid s_t,a_t)$ describes the transition dynamics of the environment,  $R(s_t,a_t)$ defines the reward associated with an action, and  $\gamma \in [0,1]$ is the discount factor, where $t$ indexes the decision steps. The Markov property assumes that the distribution of the next state depends only on the current state and action,
\begin{equation}
    P(s_{t+1}\mid s_t,a_t,s_{t-1},a_{t-1},\ldots)
    =
    P(s_{t+1}\mid s_t,a_t).
\end{equation}

The behaviour of an RL agent is described by a policy, $\pi(a\mid s)$, which defines the probability of selecting action $a$ given state $s$. In deep reinforcement learning, this policy is commonly parameterised by a neural network,
\begin{equation}
    \pi_{\theta}(a_t\mid s_t),
\end{equation}
where $\theta$ denotes the trainable network parameters. The objective is then to find a policy that maximises the expected cumulative discounted reward $J(\theta)$,
\begin{equation}
    J(\theta)
    =
    \mathbb{E}_{\pi_{\theta}}
    \left[
        \sum_{t=0}^{T-1}
        \gamma^{t} r_t
    \right],
    \label{equ:cum_dis}
\end{equation}
where $r_t$ is the reward for a given step. This approach is referred to as a policy-gradient method as the policy is being maximised via direct gradient ascent updates applied to the policy parameters $\theta$.

The discount factor, $\gamma$, controls the relative importance of immediate and future rewards. For $\gamma \rightarrow 0$, the agent becomes increasingly myopic, whereas $\gamma \rightarrow 1$ places greater importance on the long term consequences of an action. This is particularly important in sequential decision problems, where an action that appears locally optimal may lead to a poorer overall outcome. The reward function therefore plays a central role in defining the behaviour learned by the agent. It provides a numerical representation of the task objective, such that the agent is not explicitly instructed which actions to take, but instead learns behaviour
that maximises the accumulated reward.



\subsection{Actor--Critic Methods}

Policy gradient methods optimise the policy directly. Actor--critic approaches \citep{barto1983neuronlike} supplement the policy, or \emph{actor}, with a second function approximator, the \emph{critic}, which estimates the expected future return from a state,
\begin{equation}
    V_{\phi}(s_t)
    =
    \mathbb{E}_{\pi}
    \left[
        \sum_{k=0}^{T-t-1}
        \gamma^k r_{t+k}
        \,\middle|\, s_t
    \right],
\end{equation}
where $\phi$ denotes the parameters of the value network that returns the expected reward for a given state (the value function). The quality of an action relative to the expected value of the current state is described by the advantage function,
\begin{equation}
    A(s_t,a_t)
    =
    Q(s_t,a_t) - V(s_t).
\end{equation}
where the the action-value function $Q(s_t,a_t)$ is the projected reward based on a set of actions for a given state. Intuitively, $V(s)$ describes how valuable a state is, while $Q(s,a)$ describes the expected long-term return associated with taking action $a$ in state $s$. This would mean the advantage function $A$ can be interpreted as a measure of how much better or worse a specific action is compared to the average or expected action in a given state. However, the true action-value function $Q(s_t, a_t)$ is unknown, it is typically approximated by $Q(s_t, a_t) \approx r_t + \gamma V_{\phi}(s_{t+1})$. Thus, a simple one-step estimate of the advantage becomes
\begin{equation}
    \hat{A}_t
    =
    r_t + \gamma V_{\phi}(s_{t+1}) - V_{\phi}(s_t).
\end{equation}

Conceptually, the actor is updated using this term as a scaling factor. Actions that yield a positive advantage are made systematically more probable in future iterations, while actions with a negative advantage are discouraged. This adjustment establishes a direct trajectory toward maximising the expected cumulative reward (see Equation~\ref{equ:cum_dis}).


\subsection{Proximal Policy Optimisation (PPO)}

PPO is an actor--critic policy-gradient method designed to improve training stability by constraining the magnitude of policy updates \citep{schulman2017proximal}. Large policy updates can cause previously useful behaviour to be lost. PPO achieves this effect by using the probability ratio
\begin{equation}
    r_t(\theta)
    =
    \frac{
        \pi_{\theta}(a_t\mid s_t)
    }{
        \pi_{\theta_{\mathrm{old}}}(a_t\mid s_t)
    }.
\end{equation}
that asks how the new policy compares against the previous one, and then optimises the clipped objective
\begin{equation}
    L^{\mathrm{CLIP}}(\theta)
    =
    \mathbb{E}_t
    \left[
        \min
        \left(
            r_t(\theta)\hat{A}_t,\,
            \mathrm{clip}
            \left(
                r_t(\theta),
                1-\epsilon,
                1+\epsilon
            \right)
            \hat{A}_t
        \right)
    \right],
\end{equation}
where $\epsilon$ controls the maximum permitted change in the policy during an update. The clipping operation discourages updates that move the new policy too far from the policy used to collect the training trajectories.

\begin{figure*}
    \centering
    \includegraphics[width=\textwidth]{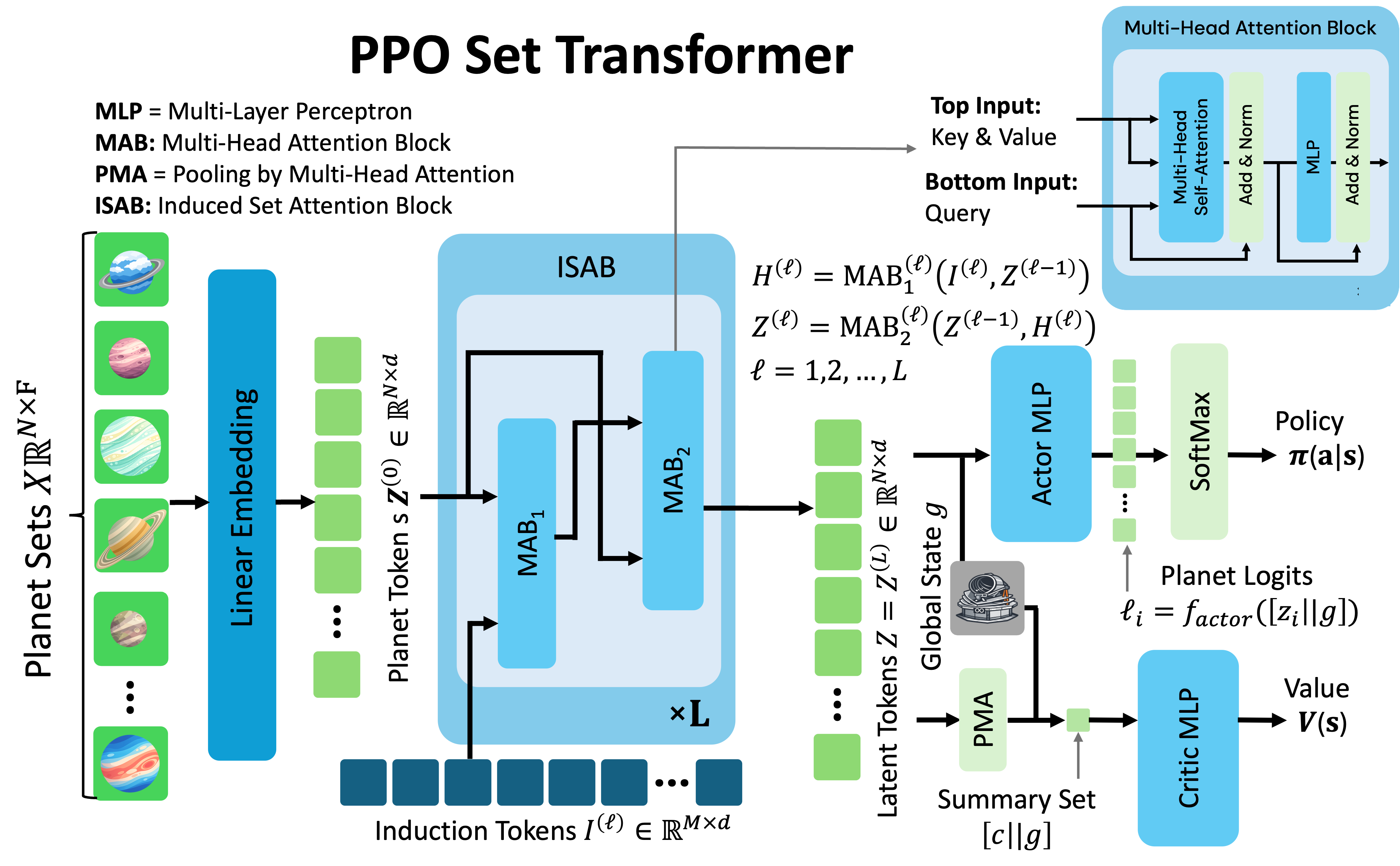}
    \caption{Set Transformer actor--critic architecture used by the RL policy. The input candidate set $X\in\mathbb{R}^{N\times F}$ is first projected into a latent $d$-dimensional representation and processed by $L$ Induced Set Attention Blocks (ISABs). Learned inducing tokens provide the queries in the first attention operation of each ISAB, after which information is redistributed to the candidate tokens. The actor combines each contextualised target representation with the global mission state to produce one logit per available observation. Each logit is then converted into a probability to select a corresponding observation. The critic uses Pooling by Multihead Attention (PMA) to obtain a permutation-invariant summary of the candidate set, which is combined with the same global state to estimate $V(s)$.}
    \label{fig:architecture}
\end{figure*}

\subsection{Training Stability}

Unlike supervised learning, the distribution of data encountered by an RL agent changes as the policy itself changes. Consequently, the training targets are non-stationary and the trajectories collected by the agent are correlated with its current behaviour. Training can therefore be sensitive to network initialisation on top of reward scaling, hyperparameters, and the degree of exploration.

This makes reinforcement learning comparatively computationally expensive, as effective policies may require large numbers of interactions with the environment and multiple training runs to establish the robustness of the resulting solution. Such constraints mean environments with expensive reward functions or evolution steps, or agent architectures with large iteration times can become computationally prohibitive for RL. However, the ability to optimise sequential decisions in large state and action spaces where individual choices alter the set and quality of decisions available later in the trajectory makes RL a natural tool for exploring telescope scheduling.

\DeclareRobustCommand{\aRieL}{aRieL} \section{\texttt{\aRieL} - RL Scheduling Environment}

The \texttt{aRieL}\footnote{\texttt{\url{https://github.com/Sympanda/aRieL}}} environment represents the evolving Ariel observing scenario. The specifics of the environment and RL design will be explained throughout this section. At each decision step $t$, the environment contains the current mission state $s_t$, from which an observation $o_t$ is constructed and passed to the agent. The agent selects a target to observe (via action $a_t$), corresponding to one of the currently available planetary systems provided in $o_t$, after which the environment executes the action and advances to the next decision point. This interaction can be summarised as
\begin{equation}
    s_t
    \rightarrow
    o_t
    \xrightarrow{\pi_\theta}
    a_t
    \rightarrow
    (r_t,s_{t+1}),
\end{equation}
where $\pi_\theta$ is the learned policy and $r_t$ is the reward associated with the resulting transition.

Following the selected action, the mission clock is advanced according to the time required to reach and complete the chosen observation. The environment then updates the observational progress of the selected target, the corresponding survey-level metrics, and the set of upcoming observation opportunities before constructing the next state. Although Ariel has a nominal four year operational lifetime, the experiments presented here adopt a 3.5 year episode, consistent with the scheduling horizon used in previous Ariel optimisation studies \citep{nakhjiri2024forgetful}.

\subsection{Target Lists} \label{sec:target list}

\begin{table}
  \centering
  \begin{tabular}{lcccccc}
    & \multicolumn{3}{c}{MCS} & \multicolumn{3}{c}{MCS + TPC sample} \\
    \cline{2-4} \cline{5-7}
    Tier & Targets & Days & \% & Targets & Days & \% \\
    \hline
    Tier~1         & 814 &  778 &  60.8 & 1427 & 1536 & 120.1 \\
    Tier~2    & 550 & 1037 &  81.1 &  817 & 1751 & 137.0 \\
    Tier~3    & 129 &   27 &   2.1 &  155 &   36 &   2.8 \\
    \hline
    Total   & 814 & 1842 & 144.1 & 1427 & 3323 & 259.9 \\
  \end{tabular}
\caption{Science time required to complete the full candidate catalogues to their available programme depths. Tier~1 gives the time required to complete all Tier~1 accessible targets, while the Tier~2 and Tier~3 rows give the additional science time required beyond the preceding tier. Percentages are relative to the adopted 3.5 year scheduling horizon. Slewing, waiting time and other operational overheads are not included in the calculated total observation time.}
\label{tab:catalogue-science-budget}
\end{table}

In this work, we use the Mission Candidate Sample from \cite{edwards2022ariel} which has been consistently updated and made publicly available on GitHub\footnote{\url{https://github.com/arielmission-space/Mission_Candidate_Sample/tree/main}}. Specifically, we use the MCS from $\text{18}^{\text{th}}$ August 2025. Given that the MCS is constantly evolving and that the number of exoplanets being detected has grown rapidly, we created an additional experiment to explore performance. We sample planets from the TESS Planetary Candidate (TPC) list as part of the same repository. We took the rate of confirmed planets from TESS\footnote{\url{https://tess.mit.edu/tess-planet-count/}} and extrapolated the rate of confirmation over four years (until the scheduled launch of Ariel). This resulted in an additional 613 planets being added to the MCS. Table \ref{tab:catalogue-science-budget} shows the breakdown of both target lists. Note, we do not consider the properties of planets when sampling from TPC. 

Even before accounting for slew or idle time, completing every baseline MCS target (excluding TPC targets) to its maximum available tier would require $144.1\%$ of the adopted scheduling horizon. The expanded catalogue increases this requirement further. The scheduling problem is therefore intrinsically oversubscribed: a policy must decide not only when to observe, but which parts of the candidate population to prioritise.

\subsection{Observation Space}
\label{sec:observations}

At each decision step, the agent receives information describing both the individual planetary systems available for observation and the current state of the mission. We therefore divide the observation into an unordered set of planet-level feature vectors and a separate mission-global vector,
\begin{equation}
    o_t =
    \left(
        \{\mathbf{x}_{i,t}\}_{i=1}^{N_t},
        \mathbf{g}_t
    \right),
\end{equation}
where $\mathbf{x}_{i,t}$ describes planetary system $i$ and $\mathbf{g}_t$ summarises the state of the observing programme. This representation avoids imposing an arbitrary ordering on the target catalogue while allowing the number and properties of available targets to change throughout an episode of the full 3.5 year mission.

Each planetary system vector contains both static and dynamic information. The static features describe properties that remain fixed throughout the mission, including the planetary and stellar properties, assigned science tier, population weighting, and preferred observing method. The dynamic features are recomputed after every action and describe the system's current observational state, including progress towards its assigned tier, the timing and contributions of the next reachable observation, and the availability of future observing opportunities. Each target representation therefore contains information describing both the scientific properties of the system and its current suitability for scheduling.

The mission-global vector contains information that cannot naturally be associated with a single target. This includes the elapsed mission time, cumulative science, slew and idle time, overall progress through the survey tiers, and the current coverage of the planetary population. These quantities provide the wider mission context required to interpret the relative value of individual observing opportunities. For example, the desirability of observing a particular system may depend not only on its immediate observability, but also on the remaining mission time and the populations that have already been sampled.

All observation features are normalised before being passed to the policy network. Planet-level features are concatenated to form a 28-dimensional vector comprising 11 static and 17 dynamic quantities, while the mission-global vector contains nine mission level quantities together with the current coverage of sufficiently populated science bins. Completed targets and padded entries are masked so that they do not contribute to subsequent attention operations. Full definitions and normalisations of all observation features are provided in Appendix~\ref{app:obs-encoding}. These features can be amended or extended as the scientific requirements of Ariel evolve.

\subsection{Set Transformer Policy}
\label{sec:set-transformer}

The planet-level observations form an unordered and dynamically changing set, making a Set Transformer \citep{lee2019set} a natural architecture for the policy. The input features for each target are first mapped to a latent token and then processed through $L$ Induced Set Attention Blocks (ISABs). Within each ISAB, a small set of learned inducing tokens first gathers information from the complete target set before redistributing this information back to the individual planet tokens. This allows each target representation to be contextualised by the competing observing opportunities available at the same decision point, while reducing the attention cost from $\mathcal{O}(N^2)$ to $\mathcal{O}(NM)$ for $M \ll N$, where N is the number of input tokens and M is the number of inducing tokens used by the ISAB architecture (see the left hand side of Figure~\ref{fig:architecture}).

After the final ISAB layer, the policy retains one contextualised latent token for each candidate target. These representations are used differently by the actor and critic. The actor must evaluate individual observing opportunities, and therefore combines each planet token with the global mission-state vector before assigning that target an action score. The critic instead evaluates the state of the survey as a whole. The complete set of planet tokens is therefore reduced to a single permutation-invariant representation using Pooling by Multi-head Attention (PMA), which is then combined with the same global mission information before estimating the state value.

Conceptually, the Set Transformer therefore allows the policy to consider each target not in isolation, but relative to the other opportunities currently available, while the separate global representation supplies the wider state of the mission. The resulting actor--critic architecture is shown in Figure~\ref{fig:architecture}, with the full Set Transformer operations given in Appendix~\ref{Appendix:set_maths}.

\subsection{Slewing}\label{sec:slew}

The telescope is represented by a single pointing $(\mathrm{RA},\mathrm{Dec})$. Choosing a target means slewing there immediately. The angular separation $\theta$ is the great-circle (haversine) distance between the current pointing and the target. We assume a constant slew rate hop
\begin{equation}
  t_{\mathrm{slew}}
  = \mathrm{clip}\bigl(\theta \times 60\,\mathrm{s\,deg}^{-1},\;
     120\,\mathrm{s},\; 7200\,\mathrm{s}\bigr),
\end{equation}
i.e.\ $1^\circ\,\mathrm{min}^{-1}$ with a 2~min settle floor and a 2~h cap. The 2~min floor is charged even for a repeat visit to the same host; hops $\gtrsim 120^\circ$ all cost 2~h. There is no acceleration profile or Earth avoidance constraint. 

The mission clock advances as slew~$\rightarrow$~idle (if early)~$\rightarrow$~observe. Arrival is $t_{\mathrm{now}}+t_{\mathrm{slew}}$. Early arrival waits until the observation block opens, and if the telescope would arrive too late for an observation, the event is not included. The pointing is updated to the selected target, so the next slew is measured from that position, with each episode initialised at the same position on the sky. Each planetary set reports this $t_{\mathrm{slew}}$ as \texttt{slew\_norm} (divided by the 2~h cap), and is paired with the first event still reachable after the slew. 

This prescription is deliberately simplified and is intended to provide a non-zero geometric cost for moving between targets rather than to reproduce the final Ariel attitude control model. The rate is a placeholder; Ariel's published science-mode performance is not used as this project aims to highlight full use of RL for Ariel scheduling rather than perfectly mimic the full observation system. Additionally, the current environment does not model detailed acceleration profiles, time dependent or field restrictions, Earth/Sun avoidance, calibration observations, or station keeping activities. These simplifications distinguish the present proof of concept environment from more operationally complete Ariel scheduling models and are discussed further in Section~\ref{sec:Dis}.

\subsection{Action Space}

Candidate actions are defined directly by the planetary systems contained within the current observation set. At each decision step, the agent is presented with a variable number of valid observing opportunities and selects one of these systems as its next action. The action space therefore changes dynamically with the observation space, with one action corresponding to each candidate system presented to the agent. The maximum set size is fixed to $N=72$, as described in Section~\ref{sec:training}, with unused entries masked when fewer candidates are available. If no targets are available, the environment continues to run until there are. The agent therefore solves a rolling local decision problem rather than receiving every target in the full catalogue as a possible action at every timestep. This restriction reduces the action space and imposes the prior that very distant future events need not be considered directly when nearer valid opportunities are available. Longer term information is instead carried through the evolving mission state and value function. We are then effectively asking the agent: given the set of exoplanets that are approaching transit and the global state of the mission, which planet is best to target for observation next. This provides further motivation for the Set Transformer policy, which naturally accommodates a variable sized set of candidate targets.

It should be noted, just like the observation space, if there are not enough targets to fill the full set transformer tokens then the actions corresponding to those tokens are also masked. 

The action space is defined directly by the candidate set presented to the agent. At decision step $t$, let
\begin{equation}
    X_t =
    \left\{
    \mathbf{x}_{t,1},
    \ldots,
    \mathbf{x}_{t,N_t}
    \right\},
    \qquad N_t \leq N,
\end{equation}
denote the set of upcoming valid observation opportunities. The corresponding action space is
\begin{equation}
    \mathcal{A}_t
    =
    \left\{
    1,\ldots,N_t
    \right\},
\end{equation}
such that selecting action $a_t=i$ schedules the observation associated with candidate $\mathbf{x}_{t,i}$.

The policy network has a maximum output dimension $N$. When fewer than $N$ candidate systems are available, the remaining positions are padding tokens and their corresponding actions are masked prior to action selection. Defining the action mask as
\begin{equation}
    m_{t,i}
    =
    \begin{cases}
        0, & i \leq N_t,\\
        -\infty, & i > N_t,
    \end{cases}
\end{equation}
the masked policy is
\begin{equation}
    \pi_\theta(a_t=i \mid o_t)
    =
    \mathrm{softmax}
    \left(
    \boldsymbol{\ell}_t + \mathbf{m}_t
    \right)_i 
\end{equation}
with $\ell$ being the corresponding logits that are coupled to the corresponding candidate-action pair. This ensures that probability mass is assigned only to observation opportunities that exist within the current candidate set.

If more than $N$ valid opportunities fall within the look-ahead window, candidates are ordered by event start time and the first $N$ are retained.

\subsection{Objectives \& Reward Function}
\label{Sect:rewards}

The reward function is designed to balance several competing objectives of the Ariel survey rather than optimise a single scheduling metric. In particular, the agent is encouraged to make progress through the observational tiers, maintain broad coverage of the target population, retain comparatively rare observing opportunities, and use the available mission time efficiently.

Scientific progress is rewarded as individual targets progress towards and complete their observational tiers. Separate weights are assigned to Tier~1, Tier~2, and Tier~3, allowing their relative scientific priority to be varied between experiments. Intermediate progress is also rewarded, with an increased contribution as a target approaches completion.

Population diversity is incorporated using planetary radius, equilibrium temperature, and host star class. Scientific progress on targets in under sampled population bins receives an increased reward, while additional rewards encourage Tier~1 coverage across the population and diversity between planetary systems. This is a simplified representation of population diversity, and the boundaries used here are specific to the present scheduling experiments rather than a final Ariel population definition. The adopted classes are given in Table~\ref{tab:population_bins}.

\begin{table}
    \centering
    \begin{tabular}{lll}
        \hline
        Axis & Class & Range \\
        \hline
        Radius & sub-Earth & $R_p < 1.5\,R_\oplus$ \\
        & super-Earth & $1.5 \le R_p < 2.5\,R_\oplus$ \\
        & mini-Neptune & $2.5 \le R_p < 4.0\,R_\oplus$ \\
        & Neptune & $4.0 \le R_p < 6.0\,R_\oplus$ \\
        & Saturn & $6.0 \le R_p < 10.0\,R_\oplus$ \\
        & Jupiter & $R_p \ge 10.0\,R_\oplus$ \\
        \hline
        Temperature & cold & $T_{\rm eq} < 400\,{\rm K}$ \\
        & warm & $400 \le T_{\rm eq} < 900\,{\rm K}$ \\
        & hot & $900 \le T_{\rm eq} < 1400\,{\rm K}$ \\
        & very hot & $1400 \le T_{\rm eq} < 2000\,{\rm K}$ \\
        & ultra hot & $T_{\rm eq} \ge 2000\,{\rm K}$ \\
        \hline
        Stellar class & M & $T_{\rm eff}<3900\,{\rm K}$ \\
        & K & $3900 \le T_{\rm eff}<5200\,{\rm K}$ \\
        & G/F & $5200 \le T_{\rm eff}<7500\,{\rm K}$ \\
        & A/F hot & $T_{\rm eff}\ge7500\,{\rm K}$ \\
        \hline
    \end{tabular}
        \caption{Population classes used by the coverage component of
    the reward. Only combinations represented in the adopted
    catalogue contribute to the population-coverage denominator.}
    \label{tab:population_bins}
\end{table}

A separate rarity component favours targets with less frequent observing opportunities, based on orbital period, while scheduling efficiency is encouraged by penalising slew and idle time and rewarding the fraction of each step spent observing. Missed observation opportunities can additionally be penalised. The reward per step can therefore be written abstractly as
\begin{equation}
\begin{aligned}
r_t ={}&
M_{\mathrm{population}}
\left(R_{\mathrm{progress}} + R_{\mathrm{tier}}\right)
+ R_{\mathrm{population}}
+ R_{\mathrm{rarity}} \\
&+ R_{\mathrm{efficiency}}
- P_{\mathrm{overhead}}
- P_{\mathrm{miss}},
\end{aligned}
\end{equation}
where $M_{\mathrm{population}}$ increases the scientific reward for targets in under sampled population bins, while $R_{\mathrm{population}}$ contains the additional population coverage and planetary system diversity rewards. Survey level milestone and terminal rewards additionally encourage progress across the complete observing programme. Each term shown in Equation~17 contains several tunable sub-components.

The relative weights of these components define the priorities presented to the agent and must therefore be chosen jointly. We do not claim that the adopted weights represent a unique or optimal Ariel objective function. Instead, they were selected to balance target completion, population diversity, rare observing opportunities, and observing efficiency. We additionally vary these weights to examine how the learned policy responds to changes in the assumed scientific priorities. The complete reward configuration and individual component weights used for each experiment are provided with the released code and configuration files.

\subsection{Training}\label{sec:training}

The PPO implementation and training pipeline were developed using Stable-Baselines3 \citep{raffin2021stable}, with a custom Set Transformer feature extractor and actor--critic policy. The final policy uses a maximum candidate-set size of $N=72$, $M=24$ inducing points per ISAB, and $L=3$ ISAB layers.

To select $N$, we inspected the number of valid observing opportunities within the adopted look-ahead window. We adopt $N=72$ as a compromise between retaining the available candidate set and limiting padding and computational cost. Having fixed $N$, the number of inducing points was tested over $M=8,12,16,20,24,28,32$ for a one year schedule. The smallest value for $M$ providing comparable performance, defined by the total reward, was retained for the final architecture. 

The final baseline policy was trained for $2,500,000$ environment interactions using episodes spanning 3.5 years. PPO hyperparameters are listed in Appendix~\ref{app:training_parameters}. During evaluation, actions are selected deterministically using the highest probability valid action. All models were trained on a MacBook Pro with an Apple M3 processor using the PyTorch MPS backend; despite the lack of dedicated high performance accelerator hardware, the models could be trained effectively on a standard consumer machine. On the same Apple M3 hardware, training the final baseline policy required approximately 15 hours. Once trained, deterministic generation of a complete 3.5-year episode required 20~s (across 200 sample runs). Only a single training seed (42) is reported in this work, and the results therefore characterise this learned policy rather than the variance of PPO training.

\section{Results}

We benchmark the RL ISAB policy against a series of baseline and heuristic policies defined below:

\begin{itemize}
    \item \textbf{Random Action:} This policy selects a random planet from the list of available exoplanets in the action space. This provides a simple baseline against which the other policies can be compared.
    
    \item \textbf{Next Available:} This policy selects the target that is next to become available for observation, regardless of its tier or system parameters. 
    
    \item \textbf{Greedy:} For a given state, this policy selects the available planet that provides the largest immediate reward.
    
    \item \textbf{Hill Climb:} This policy uses a simple set of weights to score the currently available targets and selects the target with the highest score. The weights determine how strongly different target properties influence this choice. To optimise these weights, the policy first generates a complete mission schedule and records its total reward. The weights are then slightly changed and a new complete schedule is generated. If the new schedule achieves a higher reward, the new weights are retained, otherwise the previous weights are kept. This process is repeated for 100 iterations, after which the best performing set of weights is used for evaluation. Because each trial generates a new schedule through the same environment, the Hill Climb policy is subject to the same available actions and scheduling constraints as the other policies.
\end{itemize}

\subsection{Competitive Scheduling Across Survey Scenarios}

The five policies are tested across two working scenarios described by the target lists in Section~\ref{sec:target list}. MCS\textsubscript{Full Mission} represents the current candidate sample, while MCS\textsubscript{Full Mission} $+$ TPC represents a possible future sample containing substantially more targets as further discoveries are made.

\begin{figure*}
    \centering
    \includegraphics[width=1\linewidth]{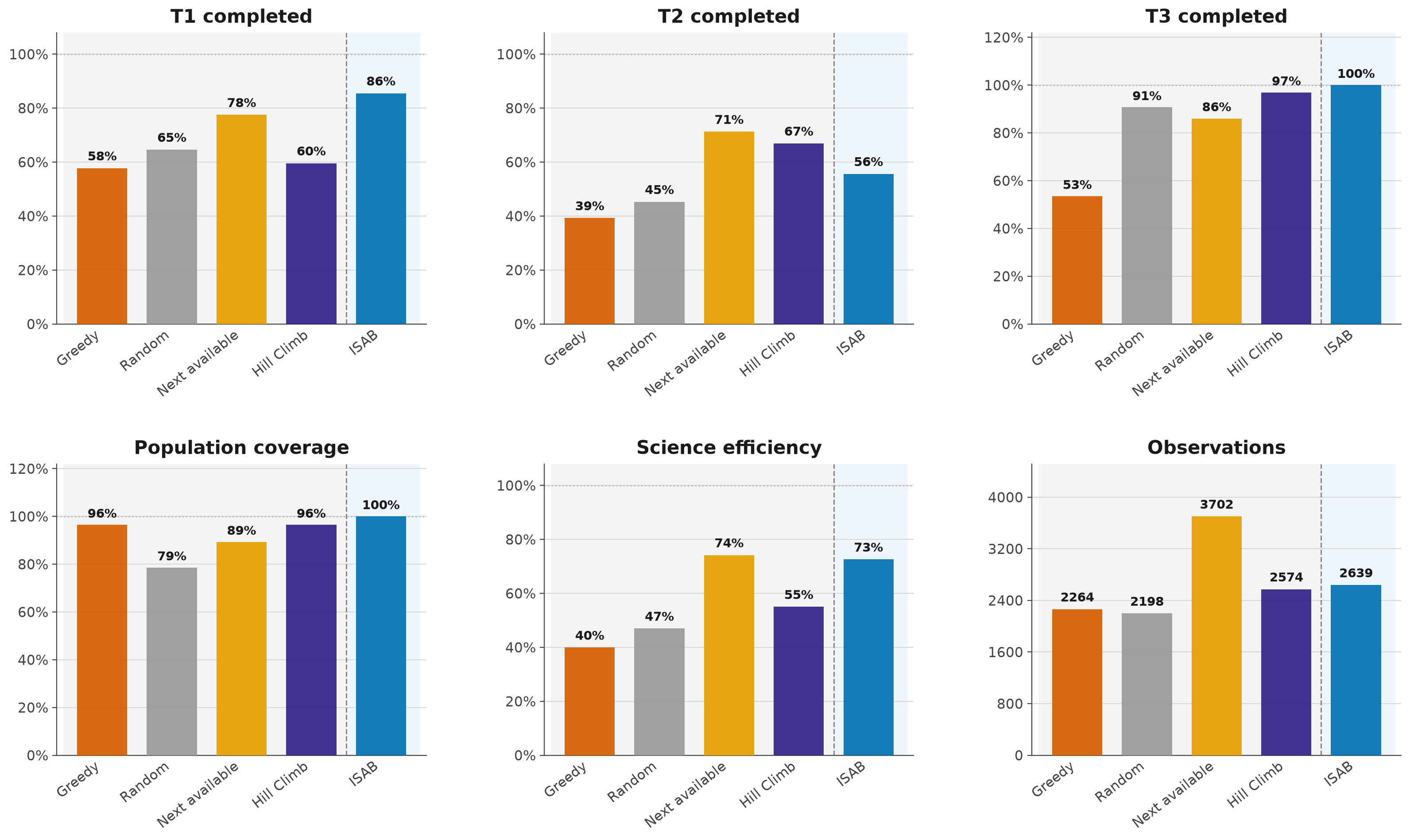}
    \caption{Visual comparison of policy performance for the baseline MCS\textsubscript{Full Mission} experiment. The panels show Tier~1, Tier~2 and Tier~3 completion, population coverage, observing efficiency, and the total number of individual observations. The observation count is included as a diagnostic of policy behaviour rather than as a direct measure of survey performance. ISAB denotes the RL Set Transformer policy. Full numerical results are given in Table~\ref{tab:exp1}, with the expanded-catalogue results reported separately in Table~\ref{tab:exp2}.}
    \label{fig:y3_comp}
\end{figure*}

Tables~\ref{tab:exp1} and \ref{tab:exp2} show the results for these two scenarios, with bold values indicating the highest value for each metric. Figure~\ref{fig:y3_comp} provides a visual representation of the principal metrics from the baseline MCS\textsubscript{Full Mission} experiment, allowing the relative behaviour of the policies to be seen more clearly. The number of individual observations is additionally shown as a diagnostic. Unlike the other quantities, this is not itself used as a measure of survey performance, but helps distinguish policies that obtain similar survey outcomes through different numbers of observing actions.

Across MCS\textsubscript{Full Mission}, the RL policy produces the largest Tier~1 sample, completing 696 targets ($85.5\%$), while also completing all 129 Tier~3 targets and achieving $100\%$ population coverage. As shown in Figure~\ref{fig:y3_comp}, this is achieved while retaining a high observing efficiency of $72.6\%$, close to the $74.2\%$ achieved by Next Available, a policy that is designed to explicitly maximise observational efficiency. The competing policies can perform better on individual metrics, with Next Available completing more Tier~2 targets and more targets overall, but doing so while achieving lower Tier~1 yield and population coverage.

The same general behaviour is retained when the candidate population is expanded in MCS\textsubscript{Full Mission} $+$ TPC. The RL policy completes 930 Tier~1 targets ($65.2\%$), compared with 730 ($51.2\%$) for the next highest policy, while maintaining $98.7\%$ Tier~3 completion. It also achieves the highest observing efficiency ($74.4\%$) and population coverage ($95.4\%$). Hill Climb achieves a higher Tier~2 completion rate ($48.3\%$ compared with $43.0\%$), while Next Available completes the largest fraction of targets overall. The relative performance shown in Figure~\ref{fig:y3_comp} therefore demonstrates that the RL policy does not maximise a single survey metric, but instead consistently finds a strong compromise across the competing survey objectives.

An additional point that is consistent across the experiments is that the Greedy policy is frequently outperformed by the Random policy. This behaviour is likely related to the interaction between the population and rarity components of the reward function and the strictly local decision making of the Greedy policy. Targets with high immediate scientific value can remain attractive even when selecting them requires substantial idle time, leading the Greedy policy to sacrifice observing throughput for locally high reward actions. Its comparatively low total observation time is consistent with this interpretation. This highlights the difficulty of navigating the multi-objective reward function using only immediate reward and provides a useful contrast with the longer horizon optimisation of the RL policy.

\subsection{Population Diversity}

Ariel aims to observe a population spanning a broad range of planetary and stellar properties rather than concentrating only on the most readily observable targets. Population coverage is therefore included directly within the reward, with a population bin considered covered once at least one target within that radius, equilibrium temperature and stellar class has completed Tier~1.

The coverage metric used in Figure~\ref{fig:y3_comp} and Tables~\ref{tab:exp1}--\ref{tab:exp4} is a summary of the population binning from the reward function described in Table~\ref{tab:population_bins}. A population bin is considered covered once at least one target within that bin has completed Tier~1. If $\mathcal{B}$ denotes the set of occupied population bins, the reported coverage is therefore
\begin{equation}
C =
\frac{1}{|\mathcal{B}|}
\sum_{b\in\mathcal{B}}
\mathbb{I}\left[N^{\rm T1}_b > 0\right],
\end{equation}
where $N^{\rm T1}_b$ is the number of Tier~1-complete targets in population bin $b$.

Within the simplified population grid adopted in this work, the RL policy maintains strong coverage while simultaneously producing the largest Tier~1 sample. For MCS\textsubscript{Full Mission}, the RL policy reaches $100\%$ coverage while completing 696 Tier~1 targets, compared with a maximum coverage of $96.4\%$ for the competing policies. When the target population is expanded, the RL policy still achieves the highest coverage at $95.4\%$, while increasing its Tier~1 sample to 930 targets.

\begin{figure*}
    \centering
    \includegraphics[width=1\linewidth]{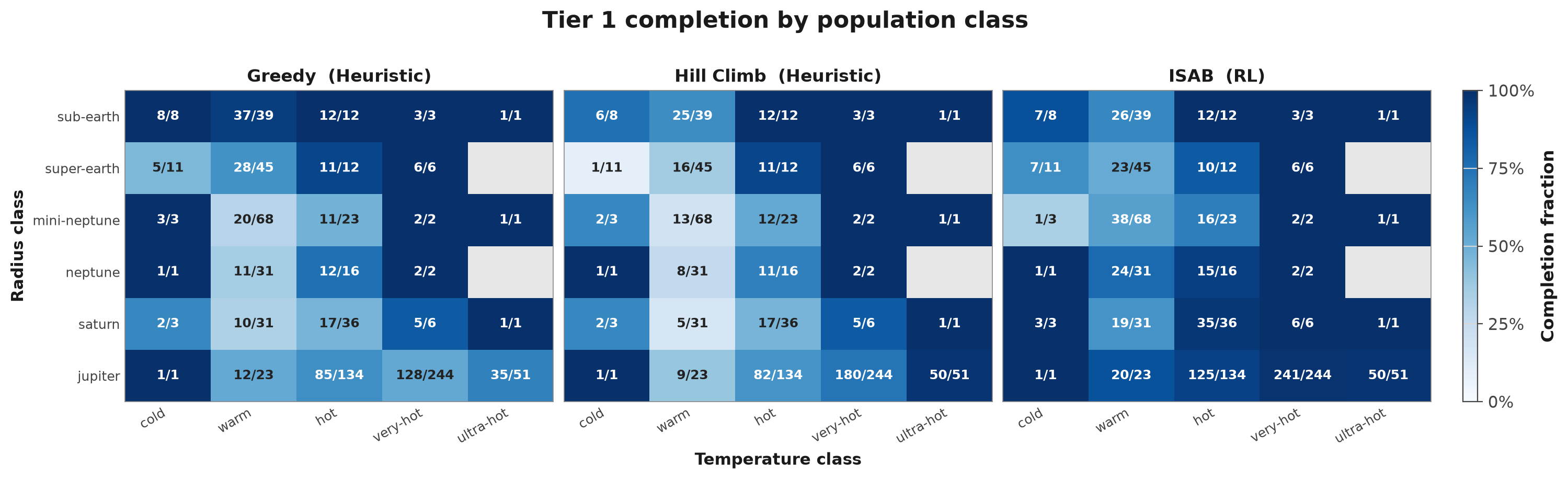}
    \caption{Tier~1 sampling across planet radius and equilibrium temperature classes after three years for the three selected policies. Each cell gives the number of targets completed to Tier~1 relative to the number available in that class, while the shading represents the corresponding completion fraction. This provides a two-dimensional projection of the population structure used by the reward, for which the full coverage metric additionally includes stellar class. A more even distribution indicates that increased Tier~1 yield has not been achieved solely by concentrating observations within the most highly populated classes.}
    \label{fig:coverage}
\end{figure*}

For clarity, Figure~\ref{fig:coverage} shows the three policies with the highest coverage metric, the RL policy, Greedy and Hill Climb optimisation baselines, providing a more detailed view of how their Tier~1 samples are distributed through the adopted population space. The RL policy does not obtain its larger Tier~1 sample simply by concentrating observations within the most highly populated regions. Instead, observations remain distributed across both common and comparatively sparsely populated classes. This allows the policy to increase the total Tier~1 yield while retaining broad coverage of the available population. For a full feature space plot, Figure~\ref{fig_app:3_feature_pop} in Figure~E1 in Appendix~\ref{app:pop_comp} additionally compares the planetary feature space sampled by the RL, Hill Climb and Greedy policies.

\subsection{Reward Adaptation}

We next test to what extent the behaviour of the RL policy can be altered through changes to the reward function. Three additional agents are trained over one year episodes to reduce the overall computational cost of this experiment, with reward structures designed to favour Tier~1 completion, Tier~2 completion, and reduced idle and slew time respectively. The baseline policies are evaluated over the same one-year horizon for comparison.

The resulting policies show clear differences in their observing strategies. Increasing the Tier~1 reward raises Tier~1 completion from $58.7\%$ for the default RL policy to $64.4\%$, but reduces Tier~3 completion from $99.2\%$ to $46.5\%$. Similarly, the Tier~2 focused policy increases Tier~2 completion from $26.5\%$ to $36.2\%$ and increases total target completion from $17.3\%$ to $24.7\%$, but reduces Tier~1 completion and population coverage to $30.8\%$ and $53.3\%$ respectively.

The policy designed to reduce idle and slew time produces a different trade-off. Its observing efficiency of $73.0\%$ is similar to the $74.0\%$ achieved by the default policy, but its Tier~1 completion increases to $65.4\%$ while Tier~3 completion falls to $65.1\%$. The effect of changing the reward is therefore not limited to the metric being directly targeted. Instead, altering the reward changes how observing time is distributed across the survey, demonstrating that the learned strategy can be steered towards different scientific priorities while also highlighting the coupling between competing objectives.

\subsection{Impact of Repeated Tier~3 Observations}

Tier~3 is intended to provide repeated observations of selected benchmark targets, including the study of temporal atmospheric variability. The exact cadence of these observations is not fixed by the scheduling model considered here. We therefore construct a deliberately demanding hypothetical scenario, MCS\textsubscript{T3 Repeats}, in which Tier~3 progress is reset on an annual basis. This allows us to examine both the survey level cost of repeated Tier~3 completion and whether the RL
policy can adapt to a modified scheduling problem without
retraining.

In MCS\textsubscript{T3 Repeats}, all Tier~3 progress and associated reward checkpoints are reset at the beginning of each year. The RL policy is evaluated directly using the policy trained on the baseline MCS\textsubscript{Full Mission} scenario, while the Hill Climb policy is refitted under the new conditions. Despite not being trained for repeated Tier~3 observations, the RL policy completes 128 of the 129 Tier~3 targets in the first year ($99.2\%$) and 127 targets in both subsequent years ($98.4\%$). The small number of missed Tier~3 targets do not show an obvious common physical characteristic, although they tend to have longer orbital periods than the wider Tier~3 sample. At the same time, it retains the largest Tier~1 sample, completing 553 targets ($67.9\%$), and achieves the highest population coverage at $89.3\%$. Its observing efficiency of $70.4\%$ is also within $0.4$ percentage points of the highest-performing Next Available policy.

Comparing MCS\textsubscript{T3 Repeats} directly with the baseline MCS\textsubscript{Full Mission} scenario shows the wider survey cost of imposing these revisits. For the RL policy, Tier~1 completion falls from $85.5\%$ to $67.9\%$, while Tier~2 completion falls from $55.6\%$ to $37.7\%$, corresponding to decreases of 17.6 and 17.9 percentage points respectively. Population coverage also decreases from $100\%$ to $89.3\%$. In contrast, observing efficiency changes only slightly, from $72.6\%$ to $70.4\%$. The repeated Tier~3 requirement therefore primarily changes how the available observing time is distributed rather than substantially reducing the fraction of the mission spent observing. Maintaining near-complete annual Tier~3 coverage comes at the cost of approximately $18$ percentage points in both Tier~1 and Tier~2 completion, which represents an important survey-level trade-off if repeated Tier~3 observations are adopted. 


\begin{table*}
  \centering
  \begin{tabular}{lccccccccccc}
    &  & \multicolumn{2}{c}{Random} & \multicolumn{2}{c}{Next Available} & \multicolumn{2}{c}{Greedy} & \multicolumn{2}{c}{Hill Climb} & \multicolumn{2}{c}{RL ISAB} \\
    \cline{3-4} \cline{5-6} \cline{7-8} \cline{9-10} \cline{11-12}
    Tier & Max & N & \% & N & \% & N & \% & N & \% & N & \% \\
    \hline
    Tier~1 & 814 & 526 & 64.6 & 631 & 77.5 & 470 & 57.7 & 485 & 59.6 & \textbf{696} & \textbf{85.5} \\
    Tier~2 & 550 & 249 & 45.3 & \textbf{392} & \textbf{71.3} & 216 & 39.3 & 368 & 66.9 & 306 & 55.6 \\
    Tier~3 & 129 & 117 & 90.7 & 111 & 86.0 & 69 & 53.5 & 125 & 96.9 & \textbf{129} & \textbf{100.0} \\
    \hline
    Total & 814 & 318 & 39.1 & \textbf{569} & \textbf{69.9} & 344 & 42.3 & 461 & 56.6 & 452 & 55.5 \\
    \hline
    Obs.\ time & -- & -- & 47.0 & -- & \textbf{74.2} & -- & 39.9 & -- & 55.0 & -- & 72.6 \\
    Coverage & -- & -- & 78.6 & -- & 89.3 & -- & 96.4 & -- & 96.4 & -- & \textbf{100.0} \\
    \hline
    Mean rank & -- & \multicolumn{2}{c}{4.00} & \multicolumn{2}{c}{2.17} & \multicolumn{2}{c}{4.42} & \multicolumn{2}{c}{2.58} & \multicolumn{2}{c}{\textbf{1.83}} \\
  \end{tabular}
\caption{\textbf{MCS\textsubscript{Full Mission}.} Performance of the five scheduling policies over the baseline full-mission target sample. $N$ gives the number of targets completed to each tier and the corresponding percentage is relative to the maximum available sample. Obs. time denotes the fraction of the mission spent within scheduled exoplanet observation blocks, irrespective of whether the corresponding target ultimately reaches Tier~1. It is therefore an internal utilisation metric and is not directly equivalent to the completed-target observing time reported by operational Ariel schedulers. Bold values indicate the highest value in each row. Mean rank gives the average rank across all reported performance metrics, where a rank of 1 denotes the best-performing policy; tied values are assigned their average rank.}  \label{tab:exp1}

  \centering
    \begin{tabular}{lccccccccccc}
&  & \multicolumn{2}{c}{Random} & \multicolumn{2}{c}{Next Available} & \multicolumn{2}{c}{Greedy} & \multicolumn{2}{c}{Hill Climb} & \multicolumn{2}{c}{RL ISAB} \\
    \cline{3-4} \cline{5-6} \cline{7-8} \cline{9-10} \cline{11-12}
    Tier & Max & N & \% & N & \% & N & \% & N & \% & N & \% \\
    \hline
    Tier~1 & 1427 & 662 & 46.4 & 730 & 51.2 & 630 & 44.1 & 480 & 33.6 & \textbf{930} & \textbf{65.2} \\
    Tier~2 & 817 & 250 & 30.6 & 322 & 39.4 & 215 & 26.3 & \textbf{395} & \textbf{48.3} & 351 & 43.0 \\
    Tier~3 & 155 & 116 & 74.8 & 77 & 49.7 & 70 & 45.2 & 152 & 98.1 & \textbf{153} & \textbf{98.7} \\
    \hline
    Total & 1427 & 338 & 23.7 & \textbf{620} & \textbf{43.4} & 423 & 29.6 & 424 & 29.7 & 474 & 33.2 \\
    \hline
    Obs.\ time & -- & -- & 53.6 & -- & 72.0 & -- & 46.2 & -- & 67.5 & -- & \textbf{74.4} \\
    Coverage & -- & -- & 72.3 & -- & 84.6 & -- & 92.3 & -- & 89.2 & -- & \textbf{95.4} \\
    \hline
    Mean rank & -- & \multicolumn{2}{c}{4.00} & \multicolumn{2}{c}{2.67} & \multicolumn{2}{c}{4.17} & \multicolumn{2}{c}{2.83} & \multicolumn{2}{c}{\textbf{1.33}} \\
  \end{tabular}
\caption{\textbf{MCS\textsubscript{Full Mission} $+$ TPC.} Performance of the five scheduling policies for the expanded target population containing additional TESS Planet Candidates. Metrics are defined as in Table~\ref{tab:exp1}, with bold values indicating the highest value in each row.}
      \label{tab:exp2}

    \centering
  \begin{tabular}{lccccccccccccccccc}
    &  & \multicolumn{2}{c}{Random} & \multicolumn{2}{c}{Next Available} & \multicolumn{2}{c}{Greedy} & \multicolumn{2}{c}{Hill Climb} & \multicolumn{2}{c}{RL Default} & \multicolumn{2}{c}{RL T1} & \multicolumn{2}{c}{RL T2} & \multicolumn{2}{c}{RL Idle} \\
    \cline{3-4} \cline{5-6} \cline{7-8} \cline{9-10} \cline{11-12} \cline{13-14} \cline{15-16} \cline{17-18}
    Tier & Max & N & \% & N & \% & N & \% & N & \% & N & \% & N & \% & N & \% & N & \% \\
    \hline
    Tier~1 & 814 & 277 & 34.0 & 281 & 34.5 & 207 & 25.4 & 264 & 32.4 & 478 & 58.7 & \textbf{524} & \textbf{64.4} & 251 & 30.8 & \textbf{532} & \textbf{65.4} \\
    Tier~2 & 550 & 91 & 16.5 & 115 & 20.9 & 61 & 11.1 & \textbf{235} & \textbf{42.7} & 146 & 26.5 & 134 & 24.4 & \textbf{199} & \textbf{36.2} & 137 & 24.9 \\
    Tier~3 & 129 & 61 & 47.3 & 36 & 27.9 & 21 & 16.3 & 57 & 44.2 & \textbf{128} & \textbf{99.2} & 60 & 46.5 & \textbf{126} & \textbf{97.7} & 84 & 65.1 \\
    \hline
    Total & 814 & 105 & 12.9 & \textbf{218} & \textbf{26.8} & 122 & 15.0 & \textbf{250} & \textbf{30.7} & 141 & 17.3 & 68 & 8.4 & 201 & 24.7 & 96 & 11.8 \\
    \hline
    Obs.\ time & -- & -- & 53.7 & -- & 66.2 & -- & 41.5 & -- & 61.2 & -- & \textbf{74.0} & -- & 65.6 & -- & 65.5 & -- & \textbf{73.0} \\
    Coverage & -- & -- & 51.7 & -- & \textbf{70.0} & -- & \textbf{71.7} & -- & 58.3 & -- & 65.0 & -- & 65.0 & -- & 53.3 & -- & 66.7 \\
    \hline
    Mean rank & -- & \multicolumn{2}{c}{6.17} & \multicolumn{2}{c}{4.00} & \multicolumn{2}{c}{6.33} & \multicolumn{2}{c}{4.33} & \multicolumn{2}{c}{\textbf{2.75}} & \multicolumn{2}{c}{4.75} & \multicolumn{2}{c}{4.33} & \multicolumn{2}{c}{\textbf{3.33}} \\
  \end{tabular}
\caption{\textbf{MCS\textsubscript{1 Year}.} One-year comparison of the baseline policies and RL policies trained using different reward priorities. RL Default uses the baseline reward structure, while RL T1, RL T2 and RL Idle increase the relative importance of Tier~1 completion, Tier~2 completion, and reducing idle and slew time respectively. Bold values indicate the highest two values in each row.}
     \label{tab:exp3}

 \centering
\begin{tabular}{lccccccccccc}
&  & \multicolumn{2}{c}{Random} & \multicolumn{2}{c}{Next Available} & \multicolumn{2}{c}{Greedy} & \multicolumn{2}{c}{Hill Climb} & \multicolumn{2}{c}{RL ISAB} \\
    \cline{3-4} \cline{5-6} \cline{7-8} \cline{9-10} \cline{11-12}
    Tier & Max & N & \% & N & \% & N & \% & N & \% & N & \% \\
    \hline
    Tier~1 & 814 & 375 & 46.1 & 432 & 53.1 & 316 & 38.8 & 262 & 32.2 & \textbf{553} & \textbf{67.9} \\
    Tier~2 & 550 & 134 & 24.4 & \textbf{233} & \textbf{42.4} & 118 & 21.4 & 212 & 38.5 & 207 & 37.7 \\
    Tier~3 (Y1) & 129 & 58 & 45.0 & 44 & 34.1 & 22 & 17.1 & 125 & 96.9 & \textbf{128} & \textbf{99.2} \\
    Tier~3 (Y2) & 129 & 58 & 45.0 & 72 & 55.8 & 33 & 25.6 & \textbf{127} & \textbf{98.4} & \textbf{127} & \textbf{98.4} \\
    Tier~3 (Y3) & 129 & 75 & 58.1 & 91 & 70.5 & 44 & 34.1 & \textbf{127} & \textbf{98.4} & \textbf{127} & \textbf{98.4} \\
    \hline
    Total & 814 & 162 & 19.9 & \textbf{358} & \textbf{44.0} & 209 & 25.7 & 235 & 28.9 & 252 & 31.0 \\
    \hline
    Obs.\ time & -- & -- & 50.2 & -- & \textbf{70.8} & -- & 41.0 & -- & 48.7 & -- & 70.4 \\
    Coverage & -- & -- & 62.5 & -- & 82.7 & -- & 85.1 & -- & 83.9 & -- & \textbf{89.3} \\
    \hline
    Mean rank & -- & \multicolumn{2}{c}{3.88} & \multicolumn{2}{c}{2.38} & \multicolumn{2}{c}{4.38} & \multicolumn{2}{c}{2.75} & \multicolumn{2}{c}{\textbf{1.63}} \\
  \end{tabular}
\caption{\textbf{MCS\textsubscript{T3 Repeats}.} Performance when Tier~3 progress is reset at the beginning of each mission year, requiring the Tier~3 sample to be revisited annually. Tier~3 (Y1--Y3) gives the number of targets completed during each yearly cycle. The RL ISAB policy is evaluated without retraining, while Hill Climb is refitted for this scenario. Bold values indicate the highest value in each row.}
  \label{tab:exp4}
  
\end{table*}

\section{Discussion}\label{sec:Dis}

This section discusses limitations and interpretations of the RL policy and its actions, together with possible extensions of the framework and simplifications of the current environment. 

\subsection{Comparison with Existing Ariel Scheduling Methods}

Previous Ariel scheduling studies have used a range of heuristic, evolutionary and metaheuristic optimisation approaches \citep{garcia2015artificial,morales2015scheduling, morales2022ariel,nakhjiri2023hybrid,nakhjiri2024forgetful}. Direct quantitative comparison between these approaches is difficult, as the studies adopt different target samples, mission dynamics, constraints and optimisation objectives. The formulation considered here differs from these previous studies, with the primary objective being to investigate whether a learned sequential policy can provide an effective and flexible approach to Ariel scheduling. One important difference is the degree of oversubscription: the target sample adopted in the previous studies requires approximately $70\%$ of the projected mission time, whereas the MCS used here requires $144\%$. However, they account for many of the observational nuances, such as taking into account calibration, station keeping and the field of view at L2. It would be possible to train a policy that is robust to such effects by including them in the environment, however, this would take more training as this would increase the state space complexity. 

Consequently, the results reported here should not be interpreted as a direct performance comparison with these schedulers. The distinction explored in this work is methodological: rather than iteratively constructing and repairing a complete mission schedule, the RL agent learns a policy that maps the current mission state and available observation opportunities directly to the next target selection. Its objective additionally incorporates population coverage and target rarity alongside tier progress and observing efficiency. A direct quantitative comparison would require all methods to be evaluated using the same target catalogue, objective function and operational constraints.



\subsection{Discounting Bias -- \texorpdfstring{$\gamma$}{gamma}}

Referring back to Equation \ref{equ:cum_dis}, the discount factor $\gamma$ determines how strongly rewards received at later timesteps contribute to the total return. In RL environments where each timestep corresponds to a fixed amount of elapsed time, this provides a consistent way of balancing immediate and longer term objectives.

For the scheduling problem considered here, however, fixed duration timesteps are not appropriate. Observation durations are target dependent, and the environment advances according to both the duration of the selected observation and the time required to complete the relevant observational tier. Consequently, successive actions can correspond to substantially different amounts of physical time. A fixed value of $\gamma$ therefore discounts rewards according to the number of decisions made rather than the actual time elapsed. This introduces a decision step dependent bias, since two trajectories spanning a similar amount of mission time can experience different levels of discounting purely because they contain different numbers of decision steps. In this setting, this can favour trajectories containing fewer, longer duration actions when their rewards are otherwise comparable.

This bias does not undermine the demonstration of RL as a suitable approach for scheduling, but it does represent a limitation of the results presented here. In particular, the learned policy may be influenced not only by the scientific reward associated with a schedule, but also by the number and duration of the actions required to construct it.

One solution would be to scale the effective discount factor with the elapsed time between decisions, for example using a duration-dependent term such as $\gamma^{\Delta t/\tau}$. However, the standard Stable-Baselines3 implementation assumes a fixed discount factor between environment steps. Implementing time dependent discounting would therefore require modification of the RL backend or a custom implementation. As the aim of this work is to demonstrate the potential of RL for this scheduling problem, such an extension is left for future work.

\subsection{Stellar Variability}

A further source of uncertainty not considered in this work is stellar variability. The impact of stellar activity on an individual observation is not necessarily known prior to observation and may reduce the quality of the resulting data. Explicitly modeling this variability would introduce additional complexity and computational cost to the environment and is beyond the scope of this work, which primarily aims to demonstrate the suitability of RL for the scheduling problem. From an RL perspective, however, this effect could be incorporated to first order without modeling the underlying stellar variability itself. Each completed observation could instead be assigned a quality factor, $q_t \in [0,1]$, which scales its contribution towards completion of the corresponding observational tier,  
\begin{equation} 
\Delta c_t = q_t \Delta c_{\mathrm{nom}}, \end{equation}  
where $\Delta c_{\mathrm{nom}}$ is the nominal progress expected from the observation. An observation expected to provide $40\%$ of the required Tier~1 signal, for example, would contribute only $20\%$ if $q_t=0.5$. The resulting completion state would then be provided to the agent at the following decision step. Importantly, the agent need not know this quality factor when selecting the observation if the variability is not known beforehand. The observation therefore remains a valid action, but its realised contribution is determined after it has been performed. In the current implementation this is equivalent to assuming $q_t=1$ for every observation. Introducing $q_t<1$ would primarily increase the number of observations required to complete some targets, delaying associated tier-completion rewards and changing their effective observational cost. If stellar activity constraints are known prior to an observation e.g. through long term photometric monitoring indicating that the star is approaching activity minima or maxima, RL offers the flexibility to prioritise or deprioritise the target accordingly. In the case of a scientifically valuable target approaching an activity minimum, the reward could be inflated to expedite its observation. Conversely, if detrimentally high activity is predicted towards activity maximum, the corresponding opportunity could simply be removed from the valid action set at this time. Such extensions would therefore allow observational uncertainty from stellar variability to be represented naturally within the existing RL framework without requiring the agent to predict the stellar behaviour itself.

\subsection{Action Analysis}

\begin{figure*}
    \centering
    \includegraphics[width=1\linewidth, trim=0cm 1.38cm 0cm 0cm, clip]{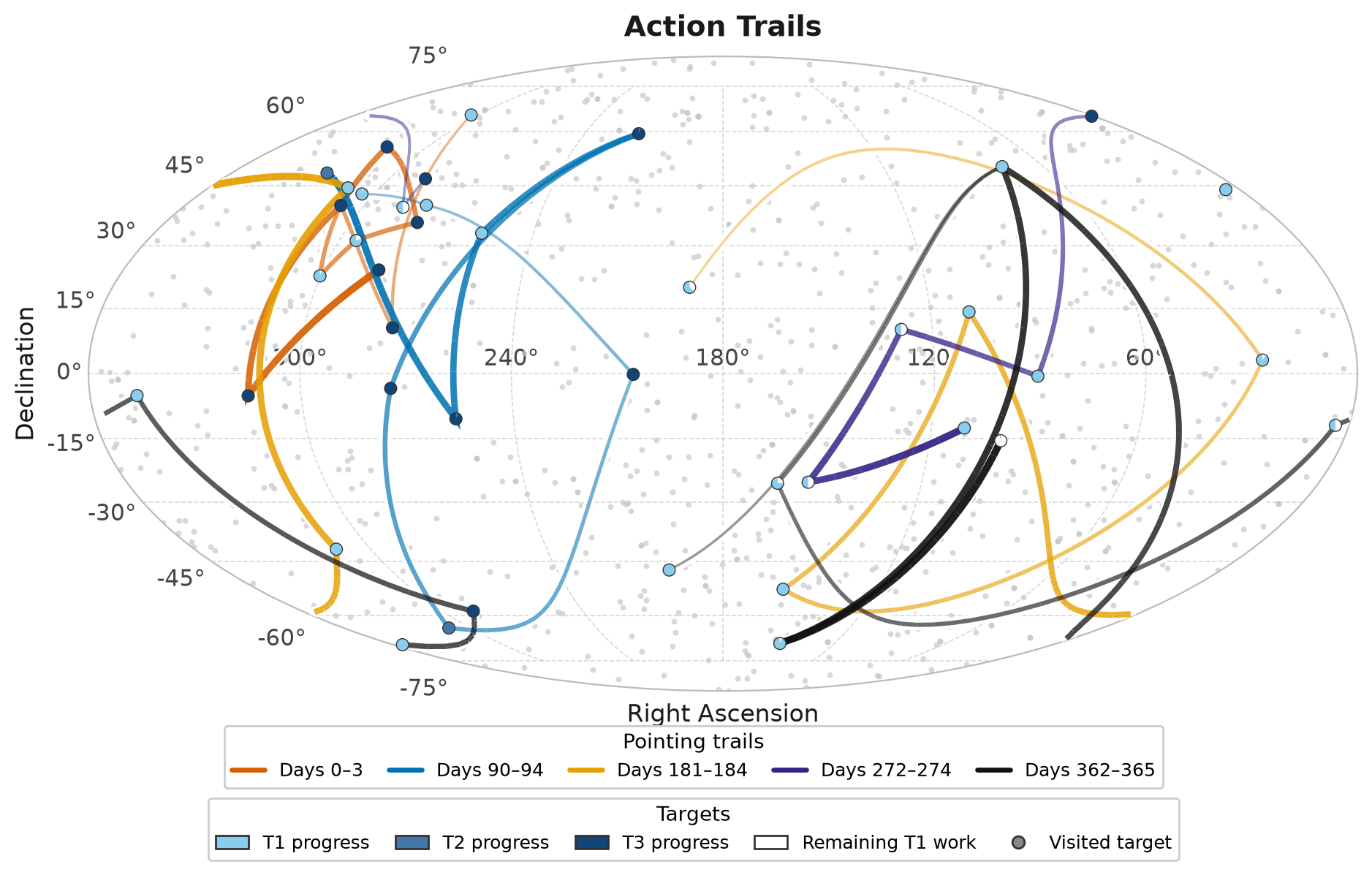}
    \includegraphics[width=1\linewidth, trim=0cm 0cm 0cm 0.7cm, clip]{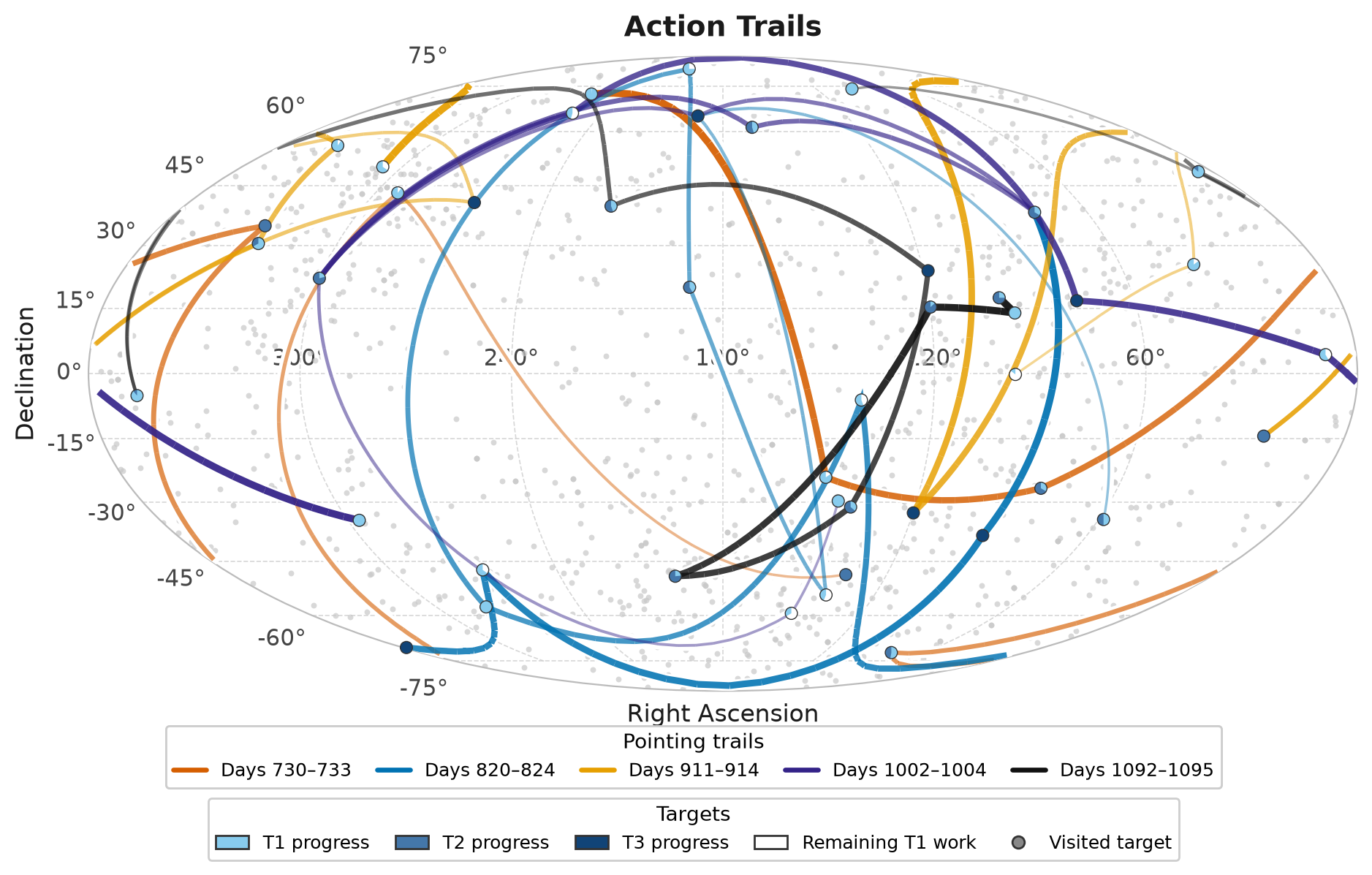}
    \caption{Action trails over the first (top) and third (bottom) year taken by the agent. Binned actions are colour coded. Thin more transparent trails indicate earlier regions in the same bin with observations being highlighted by points with small progress circles also colour coded by their tiers.}
    \label{fig:action_trails}
\end{figure*}

Over the lifetime of the observatory in this fixed episodic seed our RL agent makes 3679 observations corresponding to the same number of actions. While it's not possible to scrutinise the full action space of the model, it's possible to interpret the actions based on Figure \ref{fig:y3_comp}. Additionally, a sample of actions can be visually represented with trails of slew and target locations and progress, as shown in Figure \ref{fig:action_trails}. 

This would suggest the strategy adopted by the agent revolves around completing the Tier 3 targets relatively early (full action list in Appendix \ref{App:Actions}) as they are the highest signal to noise and yield rewards for relatively little observational time as a result. These targets can very quickly be promoted between tiers so the agent has exploited these objects as valuable and reliable observations. This is likely why most agents, unless explicitly tasked otherwise, reach very high completion levels in the first year of observing as shown in Table \ref{tab:exp3}. The exceptions being the T1-focused agent and the agent designed to minimise idle time. There does appear to be a complex relationship between minimising slew and idle time while trying to maximise the various tiered observations. For example, again in Table \ref{tab:exp3}, the Default RL policy and the Minimise Idle Time RL policy have very similar scientific efficiency (observing time) while the ratio between the T1 and T3 observations are inverted, while the agent designed to maximise T1 observation spends less total time observing. 

\subsection{Observation Completeness}

The policies differ substantially in how they distribute incomplete observations. Within the 550 Tier~2 targets, the RL policy observes every target and brings all of them to at least Tier~1, completing 306 to Tier~2 and leaving the remaining 244 at Tier~1. In contrast, Next Available and Hill Climb leave 88 and 156 Tier~2 targets unobserved, respectively, but complete Tier~2 for 392 and 368 targets.

The resulting incompleteness therefore reflects different scheduling strategies. The RL policy spends no observing time on Tier~2 targets that subsequently fail to reach Tier~1, but spends 120.9~d progressing Tier~1-complete targets that do not ultimately reach Tier~2. Other policies are more selective in which targets they begin, particularly Hill Climb, which spends only 12.3~d on such incomplete post-Tier~1 progress. This highlights a trade off between broad Tier~1 coverage of the candidate population and concentrating observations sufficiently to complete deeper tiers. 

The total share of observation time spent on incomplete targets ends up as $\sim12\%$ for the RL policy, with the Hill Climb policy being $\sim3\%$ of observation time and the remaining policies sitting at $7-15\%$. This should all be taken in the context of the total number of observations made by each policy and the total mission time spent observing.

\subsection{Stochastic Policy Sampling}

During the evaluations presented here, the RL policy is executed deterministically by selecting the valid action with the highest probability at each decision step. The policy itself is stochastic, however, and alternative valid actions can instead be sampled from $\pi_\theta(a|s)$.

Sampling in this way provides a simple means of generating multiple possible observing trajectories from the same learned policy. These trajectories could be used to examine how sensitive the resulting survey is to alternative choices in states where several actions have similar policy probabilities. The probabilities themselves should not, however, be interpreted directly as calibrated uncertainty estimates. A more complete treatment of policy uncertainty would require approaches such as independent policy ensembles or repeated training runs. 

\subsection{Integrating Target Selection with Scheduling}

Within Ariel, target selection and observation scheduling are currently treated as separate problems. In this work, the scheduling problem is considered for a fixed, predetermined MCS. However, the rapid inference of a trained RL policy provides a natural route to coupling these processes. Candidate target lists could be generated under different scientific priorities and passed directly to the scheduler, allowing their schedulability and resulting science return to be assessed under a common objective. Under this framework, target selection and scheduling could instead be optimised jointly, such that the composition of the target sample is informed by its downstream observational feasibility.

Such an approach could be valuable for exploring how alternative science requirements, tier allocations, or target selection strategies affect the achievable Ariel survey. Rather than assessing candidate samples only from their intrinsic value, the RL scheduler could provide a rapid estimate of how effectively each sample can be realised under the mission's observational constraints.

A necessary step before such an extension would be to establish that the RL policy generalises reliably across target populations that differ from those encountered during training. This would require systematic testing across perturbed or independently generated target catalogues, including changes in population balance, observation duration, visibility structure, and tier allocation. Demonstrating this robustness would allow the scheduler to act as a reusable component within future target selection studies, rather than requiring retraining for each candidate survey.




\section{Conclusions}

This work has sought to showcase how RL can be used to assist with the Ariel space telescope scheduling task, and telescope scheduling more generally. The full \texttt{aRieL} environment is publicly released alongside this paper. By combining a simulated Ariel observing environment with a Set Transformer PPO policy, we have demonstrated that long-horizon target selection can be formulated as a learnable sequential decision problem in which the agent accounts for both the properties of individual observation opportunities and the evolving state of the wider survey.

Across the deterministic mission scenarios considered, the learned policy remains competitive with, and across several objectives outperforms, the heuristic and hand-crafted baselines tested. In particular, the RL policy consistently produces large Tier~1 samples, population coverage and observing efficiency, while being near-complete for Tier~3 targets. This behaviour is retained when the target population is substantially expanded. Rather than dominating every individual metric, the RL policy generally finds a strong compromise between the competing objectives of the survey. 

A key advantage of the framework is that these objectives can be modified without changing the underlying environment or policy architecture. Altering the relative rewards associated with the observational tiers and non-observing time produces correspondingly different scheduling strategies, demonstrating that the learned policy can be steered towards different scientific priorities. These changes also expose the trade-offs between objectives, with improvements in one part of the survey generally redistributing observing time away from others. The repeated Tier~3 experiment provides a further example of this behaviour: a policy trained on the baseline mission scenario is able to generalise without retraining to annual Tier~3 revisits, still maintaining near-complete Tier~3 coverage while revealing the resulting reduction in the wider Tier~1 and Tier~2 survey. This demonstrates how the framework can be used not only to generate schedules, but also to explore the consequences of alternative observing strategies. These results motivate further exploration of reinforcement learning for astronomical telescope scheduling, with Set Transformer architectures providing a natural choice when candidate observations can be represented as unordered sets of objects of interest.

These results do not establish a globally optimal Ariel schedule, nor is the purpose of these experiments to identify a preferred Ariel reward function. Instead, they demonstrate that reinforcement learning provides a viable and flexible approach for optimising towards different scientific priorities as those priorities are specified. The Set Transformer architecture provides a natural representation for the changing set of available targets, while the RL formulation allows the longer term consequences of individual observations to influence target selection. This flexibility comes with a greater computational cost during training, and further work is required to establish generalisation across a broader range of mission futures, incorporate observational uncertainty, and compare against more extensively optimised scheduling methods. Nevertheless, the results presented here demonstrate that Ariel scheduling can be treated as a flexible, learnable optimisation problem in which the resulting observing strategy responds directly to the scientific objectives and operational constraints imposed on the agent.

\section*{Acknowledgements}

James Kostas Ray acknowledges support from the STFC UCL Centre for Doctoral Training in Data Intensive Science for Physics and Astronomy. The programme (and specifically Dr. Gabriel Facini) provided the opportunity to undertake an industry placement with Mission Zero Technologies, where the application of reinforcement learning informed the methods used in this work. A.T. acknowledges support from the ASI-INAF agreement 2021-5-HH.2-2024.

We would like to acknowledge the helpful feedback from the Ariel Consortium, which has improved the quality of this work. 

\section*{Data Availability}
 
The \texttt{aRieL} simulation environment, trained-policy configuration files, and scripts required to reproduce the experiments presented in this work are available at \texttt{\url{https://github.com/Sympanda/aRieL}}. The Ariel Mission Candidate Sample used in the experiments is derived from the publicly available Ariel Mission Candidate Sample repository, with the exact catalogue version and commit recorded in Section~\ref{sec:target list}.

\section*{AI Declaration}

Large language models (LLMs) were used during software development to assist with debugging and the generation of tests. The publicly released code was subsequently refactored with LLM assistance to improve readability, maintainability, and usability. During preparation of the manuscript, LLMs were used to assist with sentence restructuring, grammar, and the drafting and formatting of tables and equations.



\bibliographystyle{rasti}
\bibliography{example} 




\appendix

\section{Observation encoding}
\label{app:obs-encoding}

The agent observes a permutation invariant set of planet tokens together with a mission level summary vector. Each active target contributes one token of 28 features: 11 static quantities that do not change during an episode (physical properties, assigned tier, rarity weight, preferred method, and host multiplicity) and 17 dynamic quantities that are recomputed at every step (tier progress, immediate scheduling quality of the first reachable event, and a short lookahead of future opportunities). Tokens for completed or padded slots are zero and are masked out of attention. The global vector comprises nine named mission scalars (elapsed time, tier coverage, time-use breakdown, and completion counts) plus one coverage coordinate per sufficiently large population bin.

All coordinates are scaled in the environment before they enter the network. Fractions that are already in $[0,1]$ are left unchanged; remaining features are divided by a fixed physical scale (for example $20\,R_\oplus$, $3000\,\mathrm{K}$, or a $2\,\mathrm{h}$ slew cap) and then clipped, to $[-3,3]$ for planet features and $[0,1]$ for the global vector. The policy therefore receives pre normalised inputs; its first layers are linear maps of these encoded values. Tables~\ref{tab:obs-global}--\ref{tab:obs-dynamic} list every coordinate, its scale, and the range seen by the network.

\begin{table*}
\centering
\caption{Mission global observation vector. One scalar per named feature, plus one extra coordinate per population bin with at least \texttt{min\_bin\_targets} targets (default 10). All coordinates are clipped to $[0,1]$ after normalisation.}
\label{tab:obs-global}
\small
\begin{tabular}{@{}l p{5.4cm} p{4.6cm} c@{}}
\toprule
Feature & Description & Normalisation & Range \\
\midrule
\texttt{fraction\_elapsed}
  & Fraction of the mission lifetime already elapsed.
  & Already a fraction of mission length.
  & $[0,1]$ \\
\texttt{tier1\_fraction}
  & Fraction of the live catalogue that has reached Tier~1.
  & $N_{\mathrm{T1}}/N_{\mathrm{targets}}$.
  & $[0,1]$ \\
\texttt{tier2\_fraction}
  & Fraction of the live catalogue that has reached Tier~2.
  & $N_{\mathrm{T2}}/N_{\mathrm{targets}}$.
  & $[0,1]$ \\
\texttt{tier3\_fraction}
  & Fraction of the live catalogue that has reached Tier~3.
  & $N_{\mathrm{T3}}/N_{\mathrm{targets}}$.
  & $[0,1]$ \\
\texttt{used\_science\_fraction}
  & Fraction of mission time spent collecting science.
  & $T_{\mathrm{science}}/T_{\mathrm{mission}}$.
  & $[0,1]$ \\
\texttt{used\_slew\_fraction}
  & Fraction of mission time spent slewing.
  & $T_{\mathrm{slew}}/T_{\mathrm{mission}}$.
  & $[0,1]$ \\
\texttt{used\_idle\_fraction}
  & Fraction of mission time spent idle/waiting.
  & $T_{\mathrm{idle}}/T_{\mathrm{mission}}$.
  & $[0,1]$ \\
\texttt{n\_observations\_norm}
  & Cumulative number of observations executed.
  & $N_{\mathrm{obs}}/5000$.
  & $[0,1]$ \\
\texttt{n\_completed\_targets\_norm}
  & Fraction of targets that have reached \texttt{max\_tier}.
  & $N_{\mathrm{done}}/N_{\mathrm{targets}}$.
  & $[0,1]$ \\
\texttt{bin\_coverage}$_b$
  & Per-bin Tier-1 coverage (one feature per large population bin).
  & $N_{\mathrm{T1},b}/N_b$ (normalised per bin, not by catalogue size).
  & $[0,1]$ \\
\bottomrule
\end{tabular}
\end{table*}

\begin{table*}
\centering
\caption{Static per-planet features (time invariant; cached at episode reset). After normalisation the full planet vector is clipped to $[-3,3]$.}
\label{tab:obs-static}
\small
\begin{tabular}{@{}l p{5.2cm} p{4.8cm} c@{}}
\toprule
Feature & Description & Normalisation & Range \\
\midrule
\texttt{planet\_radius\_norm}
  & Planetary radius.
  & $R_p/(20\,R_\oplus)$.
  & $[0,3]$ \\
\texttt{planet\_mass\_norm}
  & Planetary mass.
  & $M_p/(4000\,M_\oplus)$.
  & $[0,3]$ \\
\texttt{planet\_temperature\_norm}
  & Equilibrium temperature.
  & $T_{\mathrm{eq}}/(3000\,\mathrm{K})$.
  & $[0,3]$ \\
\texttt{period\_norm}
  & Orbital period.
  & $P/(365.25\,\mathrm{d})$.
  & $[0,3]$ \\
\texttt{stellar\_temperature\_norm}
  & Host effective temperature.
  & $T_{\mathrm{eff}}/(10^4\,\mathrm{K})$.
  & $[0,3]$ \\
\texttt{stellar\_metallicity}
  & Host metallicity $[\mathrm{Fe/H}]$ (missing values filled with 0).
  & $\mathrm{clip}(Z,-3,3)/1.5\,\mathrm{dex}$.
  & $[-2,2]$ \\
\texttt{distance\_norm}
  & Distance to the host.
  & $d/(1000\,\mathrm{pc})$.
  & $[0,3]$ \\
\texttt{tier\_goal\_norm}
  & Highest science tier assigned to the target.
  & $\texttt{max\_tier}/3$.
  & $\{1/3,2/3,1\}$ \\
\texttt{science\_weight}
  & Inverse-frequency rarity weight (floor $0.3$).
  & Already mapped to $[\mathrm{floor},1]$.
  & $[0.3,1]$ \\
\texttt{event\_type\_binary}
  & Preferred observing method.
  & $0$ transit, $1$ eclipse, $0.5$ either.
  & $\{0,0.5,1\}$ \\
\texttt{host\_multiplicity\_norm}
  & Number of Ariel targets sharing the same host.
  & $\min(n_{\mathrm{host}},5)/5$.
  & $[0.2,1]$ \\
\bottomrule
\end{tabular}
\end{table*}

\begin{table*}
\centering
\caption{Dynamic per planet features, recomputed each step from mission progress and the first reachable event. After normalisation the full planet vector is clipped to $[-3,3]$. Missing upcoming events are filled with 0 (except $\Delta t$ slots, which saturate at the 1-year cap).}
\label{tab:obs-dynamic}
\small
\begin{tabular}{@{}l p{5.2cm} p{4.8cm} c@{}}
\toprule
Feature & Description & Normalisation & Range \\
\midrule
\multicolumn{4}{@{}l}{\textit{Progress}} \\
\texttt{obs\_completed\_norm}
  & Equivalent observations already collected.
  & $N_{\mathrm{done}}/N_{\mathrm{T3,req}}$ (per target).
  & $[0,3]$ \\
\texttt{progress\_in\_tier}
  & Progress toward the next tier boundary.
  & Already a fraction of the current tier.
  & $[0,1]$ \\
\texttt{current\_tier\_norm}
  & Current completed tier.
  & $\texttt{current\_tier}/3$.
  & $[0,1]$ \\
\texttt{obs\_remaining\_norm}
  & Equivalent observations still needed for the next tier.
  & $N_{\mathrm{rem}}/N_{\mathrm{T3,req}}$ (per target).
  & $[0,3]$ \\
\midrule
\multicolumn{4}{@{}l}{\textit{Immediate action quality}} \\
\texttt{capture\_fraction\_now}
  & Fraction of the observation block capturable if selected now.
  & Already a fraction of block duration.
  & $[0,1]$ \\
\texttt{block\_currently\_active}
  & Whether the observation block is open at the current time.
  & Binary indicator.
  & $\{0,1\}$ \\
\texttt{time\_to\_block\_end\_norm}
  & Time remaining until the block closes.
  & $(t_{\mathrm{end}}-t_{\mathrm{now}})/(10\,\mathrm{d})$.
  & $[0,3]$ \\
\texttt{idle\_if\_selected\_norm}
  & Idle wait if this target is selected now.
  & $t_{\mathrm{idle}}/(2\,\mathrm{h})$.
  & $[0,3]$ \\
\texttt{slew\_norm}
  & Slew time from the current pointing.
  & $t_{\mathrm{slew}}/(2\,\mathrm{h})$.
  & $[0,3]$ \\
\texttt{scheduling\_slack\_norm}
  & Timing margin after slew, in units of block duration
    (negative $=$ late arrival).
  & $\mathrm{clip}((t_{\mathrm{start}}-t_{\mathrm{arrive}})/T_{\mathrm{block}},-1,10)/10$.
  & $[-0.1,1]$ \\
\midrule
\multicolumn{4}{@{}l}{\textit{Future opportunities}} \\
\texttt{dt\_next\_event\_norm}
  & Time to the next event midpoint.
  & $\min(\Delta t_1,365.25\,\mathrm{d})/(365.25\,\mathrm{d})$.
  & $[0,1]$ \\
\texttt{dt\_second\_event\_norm}
  & Time to the second future event.
  & $\min(\Delta t_2,365.25\,\mathrm{d})/(365.25\,\mathrm{d})$.
  & $[0,1]$ \\
\texttt{dt\_third\_event\_norm}
  & Time to the third future event.
  & $\min(\Delta t_3,365.25\,\mathrm{d})/(365.25\,\mathrm{d})$.
  & $[0,1]$ \\
\texttt{block\_duration\_norm}
  & Duration of the next observation block ($2.5\times T_{14}$).
  & $\min(T_{\mathrm{block}},3\,\mathrm{d})/(3\,\mathrm{d})$.
  & $[0,1]$ \\
\texttt{ephemeris\_uncertainty\_norm}
  & Epoch timing uncertainty as a fraction of period.
  & $\min(\sigma_{\mathrm{epoch}}/P,1)$.
  & $[0,1]$ \\
\texttt{remaining\_opps\_mission\_norm}
  & Remaining events in the mission, relative to the catalogue total.
  & $\min(N_{\mathrm{left}}/N_{\mathrm{avail}},1)$.
  & $[0,1]$ \\
\texttt{remaining\_opps\_season\_norm}
  & Remaining events in the next 90~days.
  & $\min(N_{90\mathrm{d}},10)/10$.
  & $[0,1]$ \\
\bottomrule
\end{tabular}
\end{table*}

\section{Full Set Transformer Policy Mathematics} \label{Appendix:set_maths}

At each decision step $t$, the currently active target catalogue is represented as an unordered set of $N_t$ planets. Each planet is described by a feature vector of dimension $F$, giving the input set
\begin{equation}
X_t \in \mathbb{R}^{N_t \times F}.
\end{equation}

The raw planet features are first projected into a common latent space using a shared embedding network,
\begin{equation}
Z_t^{(0)} = \phi_{\mathrm{emb}}(X_t),
\qquad
Z_t^{(0)} \in \mathbb{R}^{N_t \times d},
\end{equation}
where each row $z_{t,i}^{(0)} \in \mathbb{R}^{d}$ corresponds to the latent representation of a single planet. Since the same embedding is applied independently to every target, no ordering of the planet catalogue is imposed. At this stage, however, each latent token primarily represents the properties of its own planet. The Set Transformer encoder is therefore used to contextualise each token with respect to the complete set of available targets.

The encoder is constructed from $L$ Induced Set Attention Blocks (ISABs). Each ISAB layer $\ell$ contains a learned set of $M$ inducing points,
\begin{equation}
I^{(\ell)} \in \mathbb{R}^{M \times d},
\qquad
\ell = 1,\ldots,L.
\end{equation}

The inducing points provide an intermediate communication bottleneck between the $N_t$ planet tokens. Each ISAB consists of two multi-head attention blocks (MABs). For a generic block
\begin{equation}
\operatorname{MAB}(A,B),
\end{equation}
the first argument provides the queries, while the second provides the keys and values,
\begin{equation}
    Q = AW_Q,
    \qquad
    K = BW_K,
    \qquad
    V = BW_V .
\end{equation}
with the attention operation given by
\begin{equation}
\operatorname{Attn}(Q,K,V)
=
\operatorname{softmax}
\left(
\frac{QK^{\top}}{\sqrt{d_h}}
\right)V.
\end{equation}

Thus, the first argument of the MAB determines which tokens are being updated, while the second argument supplies the information that those tokens attend to.

Within the first stage of ISAB layer $\ell$, the inducing points query the complete set of planet representations,

\begin{equation}
H_t^{(\ell)}
=
\operatorname{MAB}_{1}^{(\ell)}
\left(
I^{(\ell)}, Z_t^{(\ell-1)}
\right),
\qquad
H_t^{(\ell)} \in \mathbb{R}^{M \times d}.
\end{equation}

In this attention operation,
\begin{equation}
Q \leftarrow I^{(\ell)},
\qquad
K,V \leftarrow Z_t^{(\ell-1)}.
\end{equation}

The resulting $H_t^{(\ell)}$ therefore forms a compact, population-dependent representation of the complete target set.

The direction of attention is then reversed in the second stage. The planet tokens now act as queries, while the induced representation provides the keys and values,
\begin{equation}
Z_t^{(\ell)}
=
\operatorname{MAB}_{2}^{(\ell)}
\left(
Z_t^{(\ell-1)}, H_t^{(\ell)}
\right),
\qquad
Z_t^{(\ell)} \in \mathbb{R}^{N_t \times d},
\end{equation}
such that
\begin{equation}
Q \leftarrow Z_t^{(\ell-1)},
\qquad
K,V \leftarrow H_t^{(\ell)}.
\end{equation}

Hence, the two attention operations can be viewed as
\begin{equation}
Z_t^{(\ell-1)}
\;\xrightarrow{\;I^{(\ell)}\ \mathrm{queries}\ Z_t^{(\ell-1)}\;}
H_t^{(\ell)}
\;\xrightarrow{\;Z_t^{(\ell-1)}\ \mathrm{queries}\ H_t^{(\ell)}\;}
Z_t^{(\ell)}.
\end{equation}

The query and key--value roles are therefore exchanged between the two stages. First the inducing points query the planets, and then the planets query the induced representation. Importantly, the keys and values themselves are not interchanged; rather, the source of the query and the source of the key--value pairs are reversed.

After $L$ ISAB layers, the final encoder output is
\begin{equation}
Z_t
=
Z_t^{(L)}
=
\begin{bmatrix}
z_{t,1}^{\top} \\
\vdots \\
z_{t,N_t}^{\top}
\end{bmatrix}
\in
\mathbb{R}^{N_t \times d}.
\end{equation}

Each final latent vector $z_{t,i}$ therefore represents planet $i$ in the context of the complete currently available target population. These contextualised planet representations are then passed to the reinforcement learning actor--critic architecture: the actor retains the individual $z_{t,i}$ representations to assign a score to each candidate action, while the critic pools the complete set $Z_t$ into a single mission-level representation before estimating the value of the current state.

After the final ISAB layer, each candidate system is represented by a 
contextualised latent vector $\mathbf{z}_{t,i}$. These representations are 
used differently by the actor and critic to reflect their respective roles 
within PPO.

The actor retains the individual candidate representations, since each 
planetary system corresponds directly to a possible action. The global 
mission representation $\mathbf{g}_t$ is concatenated with each candidate 
token and passed through a shared actor network,

\begin{equation}
    \ell_{t,i}
    =
    f_{\mathrm{actor}}
    \left(
    [\mathbf{z}_{t,i} \Vert \mathbf{g}_t]
    \right),
\end{equation}

where $\ell_{t,i}$ is the action logit associated with candidate $i$ and 
$\Vert$ denotes concatenation. The probability of selecting each valid 
candidate is then obtained from the resulting logits,

\begin{equation}
    \pi_\theta(a_t=i \mid o_t)
    =
    \mathrm{softmax}
    \left(
    \boldsymbol{\ell}_t
    \right)_i .
\end{equation}

The critic instead estimates the value of the complete mission state and therefore requires a single representation of the candidate population. The final set of latent tokens is pooled using Pooling by Multi-head Attention (PMA),
\begin{equation}
    \mathbf{c}_t = \mathrm{PMA}(Z_t),
\end{equation}
before being combined with the same global mission information,
\begin{equation}
    V_\phi(o_t)
    =
    f_{\mathrm{critic}}
    \left(
    [\mathbf{c}_t \Vert \mathbf{g}_t]
    \right).
\end{equation}
This structure allows the actor to score individual observation opportunities while allowing the critic to evaluate their collective context and the overall state of the Ariel survey.

\section{PPO Training}\label{app:training_parameters}

\begin{table}
    \centering
    \begin{tabular}{lc}
        \hline
        Parameter & Value \\
        \hline
        Total training timesteps & $2.5 \times 10^{6}$ \\
        Parallel environments & $4$ \\
        Episode duration & $1278.375\,\mathrm{d}$ ($\approx 3.5\,\mathrm{yr}$) \\
        Maximum candidate set, $N$ & $72$ \\
        ISAB inducing points, $M$ & $24$ \\
        ISAB layers, $L$ & $3$ \\
        Latent dimension, $d$ & $128$ \\
        Attention heads & $4$ \\
        Discount factor, $\gamma$ & $0.999$ \\
        GAE parameter, $\lambda$ & $0.95$ \\
        PPO clipping parameter, $\epsilon$ & $0.2$ \\
        Learning rate & $3 \times 10^{-4}$ \\
        Rollout length ($n_{\mathrm{steps}}$) & $2048$ \\
        Batch size & $64$ \\
        PPO epochs per update & $10$ \\
        Entropy coefficient & $0.02$ \\
        Number of random seeds & $1$ (seed $42$) \\
        \hline
    \end{tabular}
    \caption{PPO and Set Transformer hyperparameters used for the final \texttt{aRieL} policy.}     \label{tab:training_params}
\end{table}

The parameters for the SB3 model used for the ISAB baseline model are shown in Table \ref{tab:training_params}.

\begin{figure*}
    \centering
    \includegraphics[width=1\linewidth]{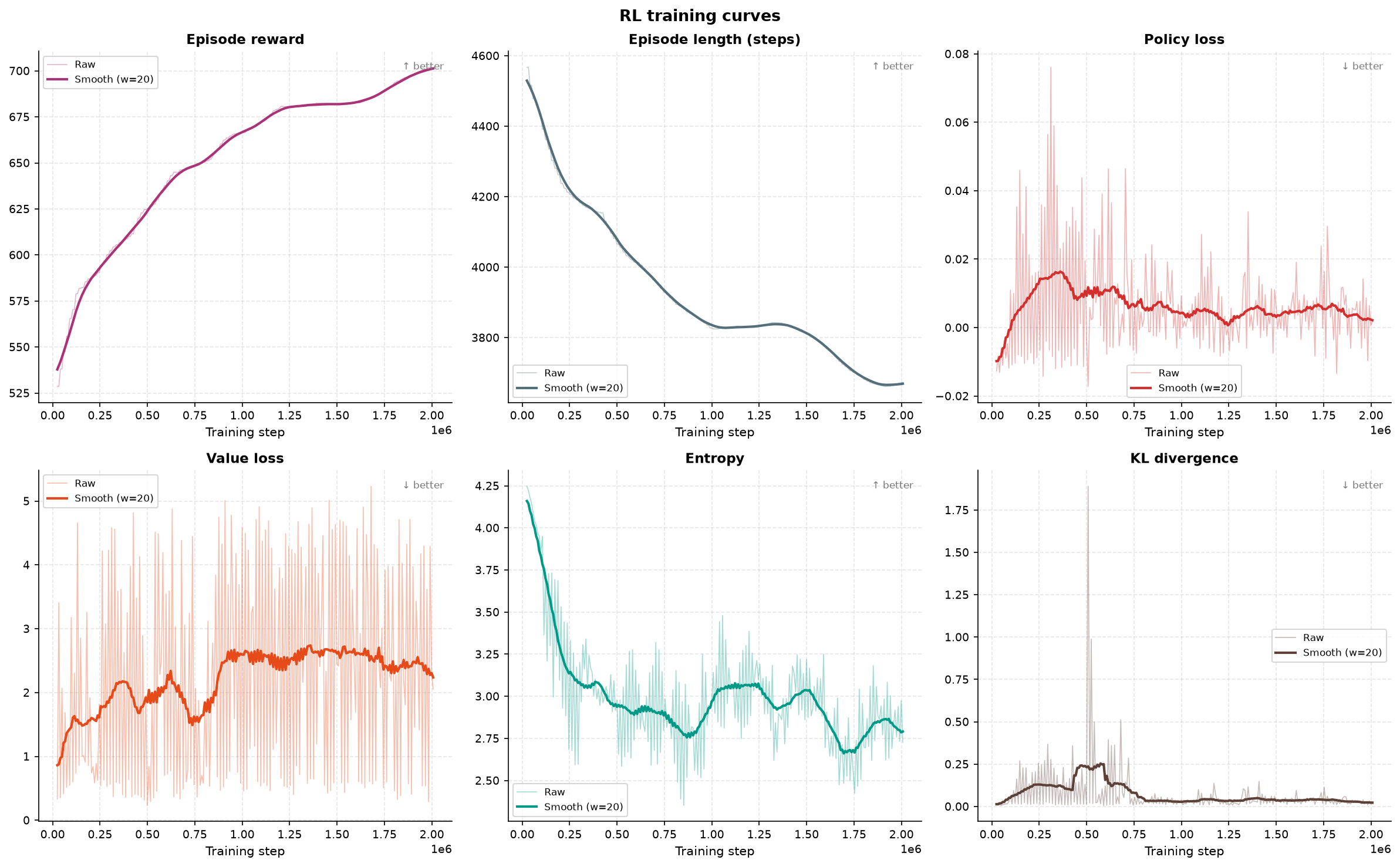}
    \caption{Training curves for PPO over 2 million training steps.}
    \label{fig:training}
\end{figure*}

Figure \ref{fig:training} shows the performance over the training of the model. The final model was trained for 2.5 million timesteps, with each episode spanning the full mission. Training used four parallel environments. The number of completed episodes remains relatively small by RL standards, and the training curves do not show clear convergence across all metrics, suggesting that additional training may continue to alter the learned policy.

\section{Full Action Recording} \label{App:Actions}

\begin{figure*}
    \centering
    \includegraphics[width=1\linewidth, trim=0cm 0cm 0cm 0.7cm, clip]{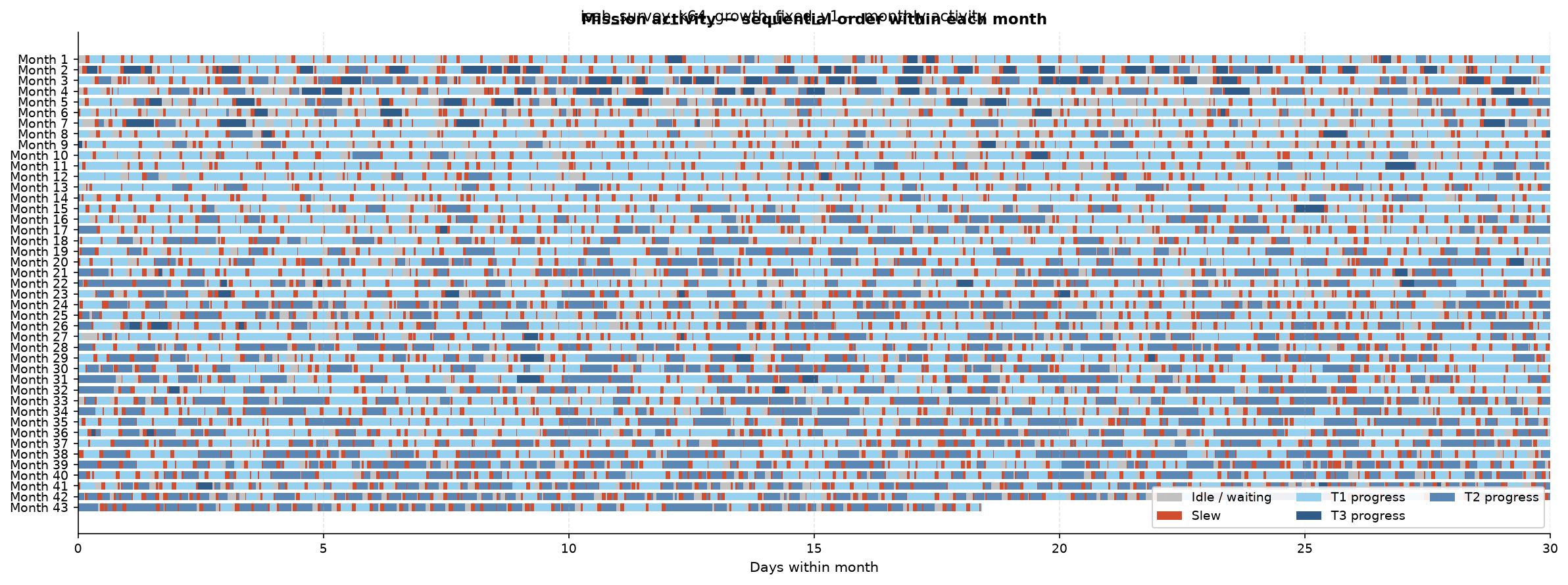}
    \caption{Visualisation of the time spent by the agent coloured coded by the type of activity for the full three and a half life cycle of the Ariel observatory. Shades of blue indicated time spent observing for respective tiers, while red indicated time slewing and grey time idle. Bars are coloured by the target’s tier at the start of each observation, so planets that finish T1 and T2 in one step never show a T2 segment and can appear as T3 work soon after.}
    \label{fig:actions_bars}
\end{figure*}

A recording of the actions and task allocation of the agent is shown here in Figure \ref{fig:actions_bars}. The time spent is colour coded by tier contributions, idle and slew time.

\section{Population Comparison}\label{app:pop_comp}

\begin{figure*}
    \includegraphics[width=1\linewidth, trim=0cm 1.5cm 0cm 0cm, clip]{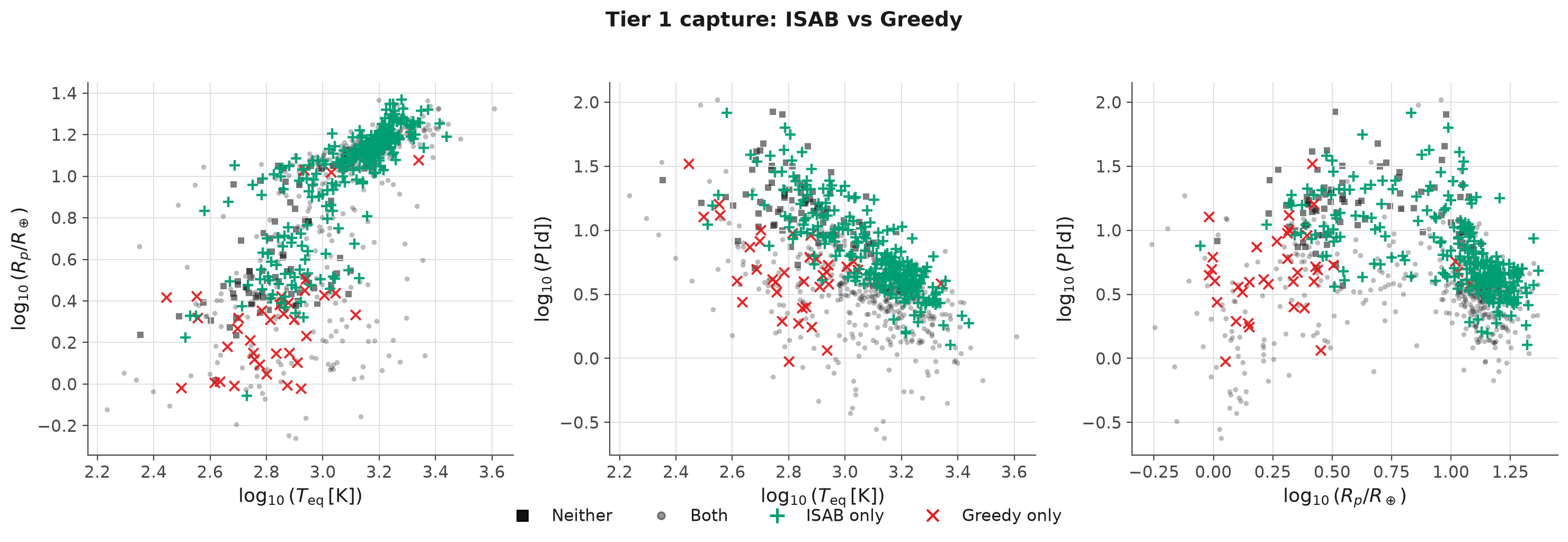}
    \includegraphics[width=1\linewidth, trim=0cm 0cm 0cm 0cm, clip]{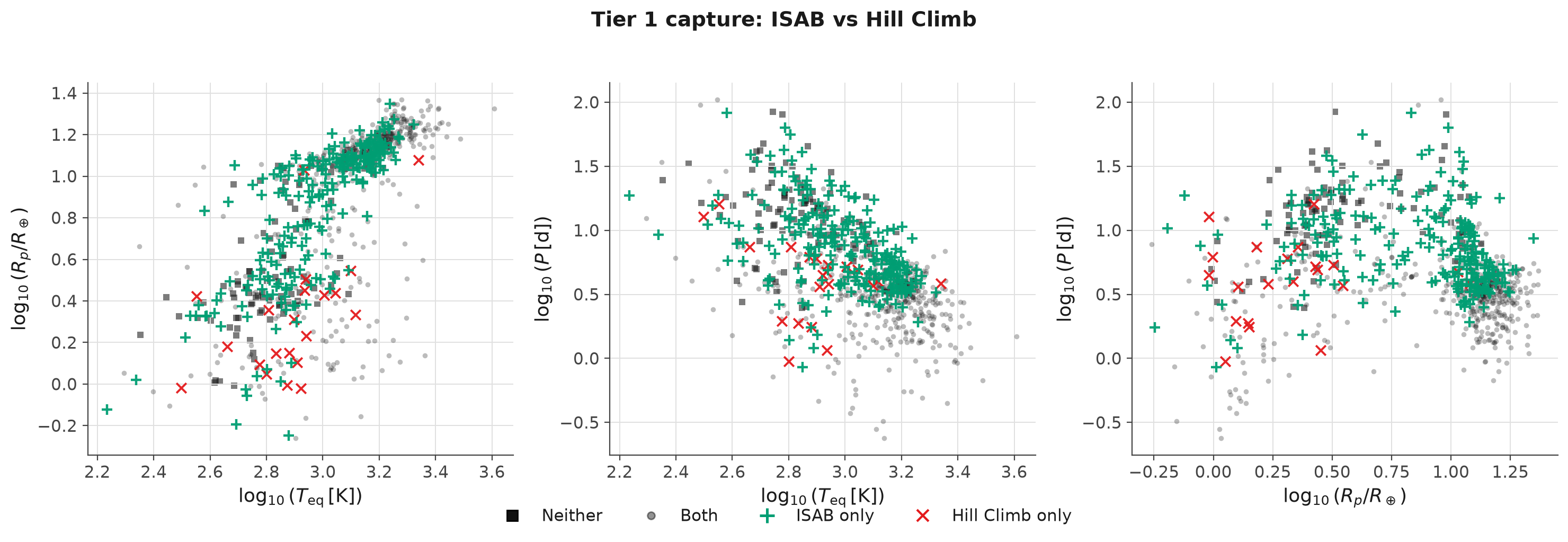}
    \caption{Pairwise planet properties for targets completed to Tier~1 by ISAB versus each baseline scheduler. Black squares mark planets missed by both; translucent black points are completed by both; green crosses are completed only by ISAB; red crosses only by the baseline.}
    \label{fig_app:3_feature_pop}
\end{figure*}

We plot the objects observed by the learned policy vs the Greedy and Hill Climb policies in the planetary feature space across temperature, radius and period in Figure~\ref{fig_app:3_feature_pop}. The main points to focus on are the objects that exist outside of the main cluster in feature space as these are key to ensure maximum coverage. An undesirable outcome would be to have isolated square points as it would mean part of feature space has been left out. For the most part, the fringes are captured by both models with some exceptions in the upper regions of period-temperature space and period-radii space. 

\bsp	
\label{lastpage}
\end{document}